\documentclass[aip,jcp,preprint]{revtex4-1}

\usepackage{amsmath,amssymb}
\usepackage[width=7in, height=9.5in]{geometry}
\usepackage{graphicx}
\usepackage{color}

\usepackage{type1cm}
\usepackage{eso-pic}
\makeatletter
\AddToShipoutPicture{%
            \setlength{\@tempdimb}{.95\paperwidth}%
            \setlength{\@tempdimc}{.48\paperheight}%
            \setlength{\unitlength}{1pt}%
            \put(\strip@pt\@tempdimb,\strip@pt\@tempdimc){%
			\makebox(0,0){\rotatebox{270}{\textcolor[gray]{0.5}%
        		{\fontsize{0.25cm}{0.25cm}\selectfont{and may be be found at \url{http://jcp.aip.org/resource/1/jcpsa6/v139/i4/p044107_s1}}}}}            
        \makebox(0,0){}
        \makebox(0,0){}                                   
        \makebox(0,0){\rotatebox{270}{\textcolor[gray]{0.5}%
        {\fontsize{0.25cm}{0.25cm}\selectfont{The following article appeared in J. Chem. Phys. 139, 044107 (2013)}}}}        
            }%
}%
\makeatother

\newcommand{\graphicsmacro}[1]{\includegraphics[width=0.48\columnwidth]{#1}}

\begin{document}

\title{Numerical Integration of the Extended Variable Generalized Langevin Equation with a Positive Prony Representable Memory Kernel}
\author{Andrew D. Baczewski}
\email{adbacze@sandia.gov}
\altaffiliation{During the preparation of this manuscript, author was also a graduate student in the Department of Electrical and Computer Engineering and the Department of Physics and Astronomy at Michigan State University, East Lansing, MI 48824.}
\author{Stephen D. Bond}
\affiliation{Multiphysics Simulation Technologies Department, Sandia National Laboratories, Albuquerque, NM 87185, USA}

\begin{abstract}
Generalized Langevin dynamics (GLD) arise in the modeling of a number of 
systems, ranging from structured fluids that exhibit a viscoelastic mechanical 
response, to biological systems, and other media that exhibit anomalous 
diffusive phenomena.  Molecular dynamics (MD) simulations that include GLD in 
conjunction with external and/or pairwise forces require the development of 
numerical integrators that are efficient, stable, and have known convergence 
properties. In this article, we derive a family of extended variable 
integrators for the Generalized Langevin equation (GLE) with a positive Prony 
series memory kernel.  Using stability and error analysis, we identify a 
superlative choice of parameters and implement the corresponding numerical 
algorithm in the LAMMPS MD software package. 
Salient features of the algorithm include exact conservation of the first and 
second moments of the equilibrium velocity distribution in some important cases, 
stable behavior in the limit of conventional Langevin dynamics, and the use of a convolution-free formalism 
that obviates the need for explicit storage of the time history of particle 
velocities.  Capability is demonstrated with respect to accuracy in numerous
canonical examples, stability in certain limits, and an exemplary application 
in which the effect of a harmonic confining potential is mapped onto a memory
kernel. \\ 
\vspace{1in} 

{\footnotesize \it \noindent Copyright 2013 American Institute of Physics. This article may be downloaded for personal use only. Any other use requires prior permission of the author and the American Institute of Physics.}
\end{abstract}

\maketitle

\section{Introduction}

Langevin dynamics\cite{coffey04} is a modeling technique, in which the motion 
of a set of massive bodies in the presence of a bath of smaller solvent 
particles is directly integrated, while the dynamics of the solvent are 
``averaged out''.  This approximation may lead to a dramatic reduction in 
computational cost compared to ``explicit solvent'' methods, since dynamics of 
the solvent particles no longer need to be fully resolved.  With Langevin 
dynamics, the effect of the solvent is reduced to an instantaneous drag force 
and a random delta-correlated force felt by the massive bodies. This framework 
has been applied in a variety of scenarios, including implicit 
solvents,\cite{uhlenbeck30,kubo66} Brownian dynamics,\cite{ermak78} dynamic 
thermostats,\cite{schneider78} and the stabilization of time 
integrators.\cite{izaguirre01}

In spite of the numerous successes of conventional Langevin dynamics it has 
long been recognized that there are physically compelling scenarios in which 
the underlying assumptions break down, necessitating a more general 
treatment.\cite{mori65,kubo66}  To this end, generalized Langevin dynamics 
(GLD) permits the modeling of systems in which the inertial gap separating the 
massive bodies from the smaller solvent particles is reduced.  Here the 
assumptions of an instantaneous drag force and a delta-correlated random force 
become insufficient, leading to the introduction of a temporally non-local 
drag and a random force with non-trivial correlations.  GLD has historically 
been applied to numerous problems over the 
years,\cite{doll76, shugard77, toxvaerd87, schweizer89} with a number of new 
applications inspiring a resurgence of interest, including 
microrheology,\cite{mason95,mason97,fricks09,didier12} biological 
systems,\cite{kou04,min05,gordon09} nuclear quantum effects,\cite{ceriotti09a,ceriotti09b,ceriotti12}
 and other situations in which anomalous diffusion arises.\cite{sokolov12}

To facilitate computational exploration of some of these applications, a 
number of authors have developed numerical integration schemes for the 
generalized Langevin equation (GLE), either in isolation or in conjunction 
with extra terms accounting for external or pairwise forces as would be 
required in a molecular dynamics (MD) simulation.  These schemes must deal 
with a number of complications if they are to remain computationally efficient 
and accurate.
\begin{itemize}
 \item The retarded drag force takes the form of a convolution of the velocity 
with a memory kernel.  This requires the storage of the velocity history, and 
the numerical evaluation of a convolution at each time step, which can become 
computationally expensive.  
 \item The generation of a random force with non-trivial correlations may also 
require the storage of a sequence of random numbers, and some additional 
computational expense incurred at each time step.
\end{itemize}
Numerous methods exist that circumvent either one or both of these 
difficulties.\cite{doll76,ciccotti80,berkowitz83,guardia85,straub86,smith90,nilsson90,wan98,gordon08}  Each has a different computational cost, implementation 
complexity, order of convergence, and specific features, e.g., some are 
restricted to single exponential memory kernels,  require linear algebra, etc.
To this end, it is difficult to distinguish any individual method as being 
optimal, especially given the broad range of applications to which GLD may be 
applied.

Motivated by the aforementioned applications and previous work in numerical 
integrators, we have developed a new family of time integration schemes for 
the GLE in the presence of conservative forces, and implemented it in a public 
domain MD code, LAMMPS.\cite{plimpton95}  Our primary impetus was to enable 
the development of reduced order models for nanocolloidal suspensions, among 
a variety of other applications outlined above.  Previous computational studies
of these systems using explicit solvents have demonstrated and resolved a 
number of associated computational challenges. \cite{intveld08}  
Our method enables a complementary framework for the modeling of these types of systems
using implicit solvents that can include memory effects.  Otherwise, to date the GLE has 
only been solved in the presence of a number of canonically tractable external potentials and memory 
kernels.\cite{kou04, desposito09, vinales06}  Integration into the LAMMPS 
framework, provides a number of capabilities. LAMMPS includes a broad array of 
external, bonded, and non-bonded potentials, yielding the possibility for the 
numerical exploration of more complex systems than have been previously 
studied.  Finally, LAMMPS provides a highly scalable parallel platform for 
studying N-body dynamics.  Consequently, extremely large sample statistics are 
readily accessible even in the case of interacting particles, for which 
parallelism is an otherwise non-trivial problem.  

In this paper, details germane to both the development of our time integration 
scheme, as well as specifics of its implementation are presented.  The time 
integration scheme to be discussed is based upon a two-parameter family of 
methods specialized to an extended variable formulation of the GLE. Some of 
the salient advantages of our formulation and the final time integration 
scheme are:
\begin{itemize}
 \item Generalizability to a wide array of memory kernels representable by a 
positive Prony series, such as power laws.
 \item Efficiency afforded by an extended variable formulation that obviates 
the explicit evaluation of time convolutions.
 \item Inexpensive treatment of correlated random forces, free of linear 
algebra, requiring only one random force evaluation per extended variable 
per timestep.
 \item Exact conservation of the first and second moments of either the 
integrated velocity distribution or position distribution, for harmonically 
confined and free particles.
 \item Numerically stable approach to the Langevin limit of the GLE.
 \item Simplicity of implementation into an existing Verlet MD framework.
\end{itemize}
The specialization to Prony-representable memory kernels is worth noting, as 
there is a growing body of literature concerning this form of the 
GLE.\cite{fricks09, mckinley09, ottobre11} A number of results have been 
presented that establish the mathematical well-posedness of this extended 
variable GLE including a term accounting for smooth conservative forces, as 
may arise in MD.\cite{ottobre11} These results include proofs of ergodicity 
and exponential convergence to a measure, as well as a discussion of the 
Langevin limit of the GLE in which the parameters of the extended system 
generate conventional Langevin dynamics.

One somewhat unique feature of our framework is that we are analyzing the GLE 
using methods from stochastic calculus.  In particular, we focus on weak 
convergence in the construction our method, i.e., error in representing the 
moments of the stationary distribution or the distribution itself.  The 
optimal parametrization of our two-parameter family of methods will be defined 
in terms of achieving accuracy with respect to this type of convergence. In 
particular, the optimal method that has been implemented achieves exactness in 
the first and second moments of the integrated velocity distribution for 
harmonically confined and free particles. Few authors have considered 
this type of analysis for even conventional Langevin integrators, with a 
notable exception being Wang and Skeel,\cite{wang03} who have carried out 
weak error analysis for a number of integrators used in conventional Langevin
dynamics.  To the best of our knowledge, this is the first time that such a weak
analysis has been carried out for a GLE integrator.  We hope that these considerations 
will contribute to a better understanding of existing and future methods.

The remainder of this paper is structured as follows:
\begin{itemize}
 \item {\bf Section \ref{problem_statement}} introduces the mathematical
details of the GLE.
 \item {\bf Section \ref{extended_variable_formalism}} presents the extended 
variable formulation and its benefits.
 \item {\bf Section \ref{numerical_integration_of_the_GLE}} develops the 
theory associated with integrating the extended variable GLE in terms of a 
two-parameter family of methods.
 \item {\bf Section \ref{error_analysis}} provides details of the error 
analysis that establishes the `optimal' method among this family.
 \item {\bf Section \ref{multistage_splitting}} discusses the extension of our method
 to a multi-stage splitting framework.
 \item {\bf Section \ref{implementation_details}} summarizes details of the 
implementation in LAMMPS.
 \item {\bf Section \ref{results}} presents a number of results that establish 
accuracy in numerous limits/scenarios, including demonstration of utility in 
constructing reduced order models.
\end{itemize}

\section{Mathematical Details}

\subsection{Statement of the Problem}\label{problem_statement}
The GLE for $N_p$ particles moving in $d$-dimensions can be written as
\begin{subequations}
\label{cont_gle}
\begin{align}
& M dV(t) = F^c\left(X(t)\right)dt - \int \limits_{0}^{t} \Gamma(t-s) V(s)ds dt + F^r(t)dt, \label{vel_gle} \\
& dX(t) = V(t)dt,
\end{align}
\end{subequations}
with initial conditions $X(0) = X_0$ and $V(0) = V_0$. Here, 
$F^c: \mathbb{R}^{dN_p} \rightarrow \mathbb{R}^{dN_p}$ is a conservative force, 
$F^r$ is a random force, $M$ is a diagonal mass matrix, and $\Gamma$ is a 
memory kernel.  The solution to this stochastic integro-differential equation 
is a trajectory, which describes the positions 
$X: \mathbb{R}^{+} \rightarrow \mathbb{R}^{dN_p}$ and velocities 
$V: \mathbb{R}^{+} \rightarrow \mathbb{R}^{dN_p}$ of the particles as a function 
of time, $t \geq 0$. The second term on the right-hand side of 
Equation~\ref{vel_gle} accounts for the temporally non-local drag force, and 
the third term accounts for the correlated random force. The nature of both 
forces are characterized by the memory kernel, 
$\Gamma: \mathbb{R}^{+} \rightarrow \mathbb{R}$, consistent with the 
Fluctuation-Dissipation theorem (FDT).\cite{kubo66,callen51}  The FDT states 
that equilibration to a temperature, $\text{T}$, requires that the two-time 
correlation of $F^r(t)$ and $\Gamma(t)$ be related as follows:
\begin{equation}
\label{fdt}
 \langle F^r_i(t+s) F^r_j(t) \rangle = \text{k}_B \text{T} \, \Gamma(s) \delta_{ij},\quad s\geq 0.
\end{equation}
Here, $\delta_{ij}$ is the Kronecker delta, and $\text{k}_B$ is Boltzmann's 
constant.

In the context of an MD simulation, we are interested in solving 
Equation~\ref{cont_gle} for both $X(t)$ and $V(t)$ at a set of $N_t$ uniformly 
spaced discrete points in time.  To this end, we seek to construct a solution 
scheme that is mindful of the following complications:
\begin{itemize}
 \item Calculation of the temporally non-local drag force requires a 
convolution with $\Gamma(t)$, and thusly the storage of some subset of the 
time history of $V(t)$.  
 \item Numerical evaluation of $F^r(t)$ requires the generation of a sequence of correlated random numbers, as specified in Equation \ref{fdt}.  
 \item As Equation~\ref{cont_gle} is a stochastic differential equation (SDE), we are not concerned 
with issues of local or global error, but rather that the integrated solution 
converges in distribution.
\end{itemize}
To circumvent the first two complications, we work with an extended variable 
formalism\cite{fricks09,mori65_2,ceriotti09b,kupferman04} in which we assume that
$\Gamma(t)$ is representable as a Prony series:
\begin{equation}
 \label{prony_memory}
 \Gamma(t) = \sum \limits_{k=1}^{N_k} \frac{c_k}{\tau_k} \text{exp}\left[-\frac{t}{\tau_k} \right], \quad t\geq 0.
\end{equation}
As will be demonstrated in Section~\ref{extended_variable_formalism}, this 
form of the memory kernel will allow us to map the non-Markovian form of the 
GLE in Equation~\ref{cont_gle} onto a higher-dimensional Markovian problem 
with $dN_k$ extended variables per particle.  The third complication is 
resolved in Sections~\ref{numerical_integration_of_the_GLE} 
and \ref{error_analysis}, in which a family of integrators is derived, and 
then `optimal' parameters are selected based upon an error analysis of the 
moments of the integrated velocity.

\subsection{Extended Variable Formalism}\label{extended_variable_formalism}

We introduce the extended variable formalism in two stages.  First, we define 
a set of extended variables that allow for an effectively convolution-free 
reformulation of Equation \ref{cont_gle}.  Then, we demonstrate that the 
non-trivial temporal correlations required of $F^r(t)$ can be effected through
coupling to an auxiliary set of Ornstein-Uhlenbeck (OU) processes.

We begin by defining the extended variable, $Z_{i,k}(t)$, associated with the 
$k$th Prony mode's action on the $i$th component of $X(t)$ and $V(t)$:
\begin{equation}
 \label{ext_var_def}
 Z_{i,k}(t) = -\int\limits_{0}^{t} \frac{c_k}{\tau_k} \exp\left[-\frac{(t-s)}{\tau_k}\right] V_i(s) ds
\end{equation}
Component-wise, Equation \ref{cont_gle} can now be rewritten as:
\begin{subequations}
\label{ext_gle_1}
\begin{align} 
 &m_idV_i(t) = F^c_i\left(X(t)\right)dt + \sum \limits_{k=1}^{N_k} Z_{i,k}(t)dt + F^r_i(t)dt, \\
 &dX_i(t) = V_i(t)dt.
\end{align}                                                                               
\end{subequations}
Rather than relying upon the integral form of Equation \ref{ext_var_def} to 
update the value of $Z_{i,k}(t)$, we consider the total differential of 
$Z_{i,k}(t)$ to generate an equation of motion that takes the form of a simple 
SDE:
\begin{equation}
 \label{ext_var_eom}
 dZ_{i,k}(t) = -\frac{1}{\tau_k} Z_{i,k}(t)dt - \frac{c_k}{\tau_k} V_{i}(t)dt
\end{equation}
Now, the system of Equations \ref{ext_gle_1} and \ref{ext_var_eom} can be 
resolved for $X_i(t)$, $V_i(t)$, and $Z_{i,k}(t)$ without requiring the 
explicit evaluation of a convolution integral.

Next, we seek a means of constructing random forces that obey the FDT, as in 
Equation \ref{fdt}.  To this end, we consider the following SDE:
\begin{equation}
 dF_{i,k}(t) = -\frac{1}{\tau_k} F_{i,k}(t)dt + \frac{1}{\tau_k}\sqrt{2\text{k}_B\text{T} c_k} dW_{i,k}(t)
\end{equation} 
If $W_{i,k}$ is a standard Wiener process, this SDE defines an 
Ornstein-Uhlenbeck (OU) process, $F_{i,k}(t)$. Using established properties of 
the OU process,\cite{kloeden_book} we can see that $F_{i,k}(t)$ has mean zero 
and two-time correlation:
\begin{equation}
 \langle F_{i,k}(t+s) F_{i,k}(t) \rangle = \text{k}_B \text{T} \frac{c_k}{\tau_k} \exp\left[-\frac{1}{\tau_k}s \right], \quad s\geq 0
\end{equation}
It is then clear, that the random force in Equation \ref{cont_gle} can be 
rewritten as:
\begin{equation}
 F^r_i(t) = \sum \limits_{k=1}^{N_k} F_{i,k}(t)
\end{equation}
Here each individual contribution is generated by a standard OU process, the 
discrete-time version of which is the AR(1) process.  While we are still 
essentially forced to generate a sequence of correlated random numbers, 
mapping onto a set of AR(1) processes has the advantage of requiring the 
retention of but a single prior value in generating each subsequent value. 
Further, standard Gaussian random number generators can be employed.  

Combining both results, the final extended variable GLE can be expressed in 
terms of the composite variable, 
$S_{i,k}(t)=Z_{i,k}(t) + F_{i,k}(t)$:
\begin{subequations}
\label{ext_gle_2}
\begin{align} 
 &m_idV_i(t) = F^c_i\left(X(t)\right)dt + \sum \limits_{k=1}^{N_k} S_{i,k}dt \\
 &dX_i(t) = V_i(t)dt \\
 &dS_{i,k}(t) = -\frac{1}{\tau_k} S_{i,k}(t)dt - \frac{c_k}{\tau_k} V_{i}(t)dt + \frac{1}{\tau_k}\sqrt{2\text{k}_B\text{T}c_k} dW_{i,k}(t)                    
\end{align}
\end{subequations}
It is for this system of equations that we will construct a numerical 
integration scheme in Section~\ref{numerical_integration_of_the_GLE}.    

It is worth noting that other authors have rigorously shown that this 
extended variable form of the GLE converges to the Langevin equation in the 
limit of small $\tau_k$.\cite{ottobre11} Informally, this can be seen by 
multplying the $S_{i,k}$ equation by $\tau_k$, and taking the limit as $\tau_k$ 
goes to zero, which results in 
\[ S_{i,k}(t)dt = - c_k V_{i}(t)dt + \sqrt{2\text{k}_B\text{T}c_k} dW_{i,k}(t). \]
Inserting this expression into the equation for $V_i$, we obtain
\begin{eqnarray*}
m_idV_i(t) &=& F^c_i\left(X(t)\right)dt - \left(\sum_{k=1}^{N_k} c_k \right) V_{i}(t)dt + \sqrt{2\text{k}_B\text{T}} \sum_{k=1}^{N_k} \sqrt{c_k} dW_{i,k}(t),\\
dX_i(t) &=& V_i(t)dt,
\end{eqnarray*}
which is a conventional Langevin equation. We have been careful to preserve 
this limit in our numerical integration scheme, and will explicitly 
demonstrate this theoretically and numerically.

\subsection{Numerical Integration of the Extended GLE}\label{numerical_integration_of_the_GLE}

We consider a family of numerical integration schemes for the system in 
Equation \ref{ext_gle_2} assuming a uniform timestep, $\Delta t$.  Notation is 
adopted such that $X_i(n\Delta t)=X^n_i$ for $n \in \mathbb{N}$.  Given the values 
of $X_i^n$, $V_i^n$, and $S^n_{i,k}$, we update to the $(n+1)$th time step using the following splitting method:
\begin{enumerate}
 \item {\it Advance $V_i$ by a half step:}
 \begin{equation}
  V^{n+1/2}_i = V^{n}_i + \frac{\Delta t}{2 m_i} F^c_i\left(X^n\right) + \frac{\Delta t}{2 m_i} \sum \limits_{k=1}^{N_k} S^n_{i,k} 
 \end{equation}
 \item {\it Advance $X_i$ by a full step:}
 \begin{equation}
  X^{n+1}_i = X^{n}_i + \Delta t V^{n+1/2}_i
 \end{equation}
 \item {\it Advance $S_{i,k}$ by a full step:}
 \begin{equation}
  \label{s_update}
  S^{n+1}_{i,k} = \theta_{k} S^{n}_{i,k} - \left(1-\theta_{k}\right) c_k V^{n+1/2}_i + \alpha_{k} \sqrt{2 \text{k}_B\text{T} c_k } B^n_{i,k}
 \end{equation}                             
 \item {\it Advance $V_i$ by a half step:}
 \begin{equation}
  V^{n+1}_i = V^{n+1/2}_i + \frac{\Delta t}{2 m_i} F^c_i\left(X^{n+1}\right) + \frac{\Delta t}{2 m_i} \sum \limits_{k=1}^{N_k} S^{n+1}_{i,k} 
 \end{equation} 
\end{enumerate}
Here, each $B^{n}_{i,k}$ is drawn from an independent Gaussian distribution of 
mean zero and variance unity. The real-valued $\theta_{k}$ and $\alpha_{k}$ can 
be varied to obtain different methods.  For consistency, we require that
\[ \theta_k = 1 - \frac{\Delta t}{\tau_k} + \mathcal{O}(\Delta t^2), \quad
\text{and} \quad 
\alpha_k = \frac{\sqrt{\Delta t}}{\tau_{k}} + \mathcal{O}(\Delta t) .\]

For the remainder of this article, we restrict our attention three different 
methods, each of which corresponds to a different choice for
$\theta_{k}$ and $\alpha_{k}$.
\begin{itemize}
 \item {\bf Method 1:}
Using the Euler-Maruyama scheme to update $S_{i,k}$ is equivalent to using
\[ \theta_{k} := 1 - \frac{\Delta t}{\tau_k} \quad \text{and} \quad
\alpha_{k} := \frac{\sqrt{\Delta t}}{\tau_{k}} \]

 \item {\bf Method 2:} 
If $V_i$ is held constant, the equation for $S_{i,k}$ can be solved exactly. 
Using this approach is equivalent to setting
\[ \theta_{k} := \exp( - \Delta t / \tau_k ) \quad \text{and} \quad
\alpha_{k} := \sqrt{\frac{(1 - \theta_k^2)}{2 \tau_k}} \]
 \item {\bf Method 3:}
Both methods 1 and 2 are unstable as $\tau_k$ goes to zero.  To improve the 
stability when $\tau_k$ is small, we consider the following modified version 
of method 2:
\[ \theta_{k} := \exp( - \Delta t / \tau_k ) \quad \text{and} \quad
\alpha_{k} := \sqrt{\frac{(1 - \theta_k)^2}{\Delta t}} \]
\end{itemize}

Note that all three methods satisfy the consistency condition, and are 
equivalent to the St\"ormer-Verlet-leapfrog method\cite{hairer03} when
$c_k = 0$ and $S_{i,k}(0) = 0$.

\subsection{Error and Stability Analysis}\label{error_analysis}
To help guide our choice of method, we compute the moments of the stationary 
distribution for a one-dimensional harmonic potential (natural frequency 
$\omega$) and a single mode memory kernel (weight $c$ and time scale $\tau$).
A similar approach has been used for the classical Langevin
equation.\cite{wang03,leimkuhler13}  
The extended variable GLE for this system converges to a distribution of the 
form
\begin{equation*}
 \rho(X,V,S) = \frac{1}{Z}\exp\left[-(mV^2+m\omega^2 X^2 + \frac{\tau}{c} S^2)/2\text{k}_B\text{T}\right]
\end{equation*} 
Where $Z$ is the usual normalization constant.  From this, we can derive the 
analytic first and second moments
\[
\langle V \rangle = 0, \quad
\langle X \rangle = 0, \quad \text{and} \quad
\langle S \rangle = 0.
\]
\[
\langle X V \rangle = 0, \quad
\langle V S \rangle = 0, \quad \text{and} \quad
\langle X S \rangle = 0.
\]
\[
\langle V^2 \rangle = \frac{\text{k}_B\text{T}}{m}, \quad
\langle X^2 \rangle = \frac{\text{k}_B\text{T}}{m\omega^2}, \quad \text{and} \quad
\langle S^2 \rangle = \frac{c\text{k}_B\text{T}}{\tau}.
\]

Next, we consider the discrete-time process generated by our numerical 
integrators, and show that the moments of its stationary distribution converge 
to the analytic ones in $\Delta t$.
\begin{eqnarray*}
V^{n+1/2} &=& V^{n} - \frac{\Delta t}{2} \omega^2 X^n + \frac{\Delta t}{2 m} S^n \\ 
X^{n+1} &=& X^{n} + \Delta t \, V^{n+1/2} \\ 
S^{n+1} &=& \theta \, S^{n} - (1 - \theta) c \, V^{n+1/2} + \alpha \sqrt{2 \text{k}_B\text{T} c } \, B^{n} \\ 
V^{n+1} &=& V^{n+1/2} - \frac{\Delta t}{2} \omega^2 X^{n+1} + \frac{\Delta t}{2 m} S^{n+1}
\end{eqnarray*}
The stationary distribution of this process is defined by the time 
independence of its first moments
\[
\langle X^{n+1} \rangle = \langle X^n \rangle, \quad
\langle V^{n+1} \rangle = \langle V^n \rangle, \quad \text{and} \quad
\langle S^{n+1} \rangle = \langle S^n \rangle
\]
Enforcing these identities, it can be shown that
\[
\langle V^{n} \rangle = 0, \quad
\langle X^{n} \rangle = 0, \quad \text{and} \quad
\langle S^{n} \rangle = 0.
\]
Hence, the first moments are correctly computed by the numerical method for 
any choice of $\theta$ and $\alpha$. Computing the second moments, we obtain
\[
\langle X^n V^n \rangle = 0, \quad
\langle V^n S^n \rangle = 0, \quad 
\langle X^n S^n \rangle = \frac{2 \Delta t^2 \alpha^2 c \, \text{k}_B\text{T} }
{2 c \Delta t (1 - \theta)^2 - m (1 - \theta^2) (4 - (\omega \Delta t)^2)},
\]
\[
\langle (V^{n})^2 \rangle = \frac{\text{k}_B\text{T} }{m} \frac{\Delta t \alpha^2}{(1 - \theta)^2}, \quad
\langle (X^{n})^2 \rangle =  \frac{\text{k}_B\text{T} }{m \omega^2} \frac{\Delta t \alpha^2 ( 2 c \Delta t (1 - \theta)^2 - 4 m (1 - \theta^2) )}{(1 - \theta)^2 (2 c \Delta t (1 - \theta)^2 - m (1 - \theta^2)(4 - (\omega \Delta t)^2))}, \]
\[ \text{and} \quad
\langle (S^{n})^2 \rangle = \frac{c \, \text{k}_B\text{T}}{\tau} \frac{ 2 \tau \alpha^2 m (4 - (\omega \Delta t)^2)}{m (1 - \theta^2)(4 - (\omega \Delta t)^2) - 2 c \Delta t (1 - \theta)^2 }
\]
From this analysis, we conclude that we obtain the correct second moment for 
$V$ for any method with $\Delta t \alpha^2 = (1 - \theta)^2$.  Now, applying 
the particular values of $\theta$ and $\alpha$, and expanding in powers of 
$\Delta t$, we obtain the following.
\begin{itemize}
 \item {\bf Method 1:}
\[
\langle (V^{n})^2 \rangle = \frac{\text{k}_B \text{T}}{m} , \quad
\langle (X^{n})^2 \rangle = \frac{\text{k}_B \text{T}}{m \omega^2}
\left( 1 + \frac{(\omega \Delta t)^2}{4}  \right) + \mathcal{O}(\Delta t^4), \]
\[ 
\langle X^n S^n \rangle = \frac{\Delta t^2 \, c \, \text{k}_B \text{T}}{4 m \tau} + \mathcal{O}(\Delta t^3),
\quad \text{and} \quad 
\langle (S^{n})^2 \rangle = \frac{c \, \text{k}_B\text{T}}{\tau}
\left( 1 + \frac{\Delta t}{2 \tau}  \right) + \mathcal{O}(\Delta t^2)
\]
\item {\bf Method 2:}
\[
\langle (V^{n})^2 \rangle = \frac{\text{k}_B\text{T}}{m}
\left(1 + \frac{\Delta t^2}{12 \tau^2}  \right)
+ \mathcal{O}(\Delta t^4), \quad
\langle (X^{n})^2 \rangle = \frac{\text{k}_B\text{T}}{m \omega^2}
\left(1 + \frac{(1 + 3 (\omega \tau)^2)\Delta t^2}{12 \tau^2} \right) + \mathcal{O}(\Delta t^4),\]
\[ 
\langle X^n S^n \rangle = \frac{\Delta t^2 \, c \, \text{k}_B \text{T}}{4 m \tau} + \mathcal{O}(\Delta t^4),
\quad \text{and} \quad 
\langle (S^{n})^2 \rangle = \frac{c \, \text{k}_B\text{T}}{\tau}
\left(1 + \frac{c \Delta t^2}{4 m \tau}  \right) 
+ \mathcal{O}(\Delta t^4)
\]
 \item {\bf Method 3:}
\[
\langle (V^{n})^2 \rangle = \frac{\text{k}_B \text{T}}{m} , \quad
\langle (X^{n})^2 \rangle = \frac{\text{k}_B \text{T}}{m \omega^2}
\left( 1 + \frac{(\omega \Delta t)^2}{4}  \right) + \mathcal{O}(\Delta t^4), \]
\[
\langle X^n S^n \rangle = \frac{\Delta t^2 \, c \, \text{k}_B \text{T}}{4 m \tau} + \mathcal{O}(\Delta t^4),
\quad \text{and} \quad 
\langle (S^{n})^2 \rangle = \frac{c \, \text{k}_B\text{T}}{\tau}
\left( 1 + \frac{(3 c \tau - m)\Delta t^2}{12 m \tau^2}  \right) + \mathcal{O}(\Delta t^4)
\]
\end{itemize}
For methods 1 and 3, we obtain the exact variance for $V$, independent of 
$\Delta t$, since they both satisfy $\Delta t \alpha^2 = (1 - \theta)^2$. For 
method 2, the error in the variance of $V$ is second-order in $\Delta t$.  All 
three methods overestimate the variance of $X$, with an error which is 
second-order in $\Delta t$.  The error in the variance of $S$ is first-order 
for method 1, and second-order for methods 2 and 3.

It is possible to choose $\theta$ and $\alpha$ to obtain the exact variance 
for $X$, but this would require using a different value for $\theta$ and 
$\alpha$ for each value of $\omega$.  This is not useful in our framework, 
since the method is applied to problems with general nonlinear interaction 
forces.

We would like our numerical method to be stable for a wide range of values for 
$\tau_k$. As we mentioned in Section~\ref{extended_variable_formalism}, the 
GLE converges to the conventional Langevin equation as $\tau_k$ goes to zero, 
and we would like our numerical method to have a similar property.  For fixed
$\Delta t$, both methods 1 and 2 are unstable ($\alpha_k$ is unbounded) as 
$\tau_k$ goes to zero.  However, method 3 does not suffer from the same 
problem, with $\theta_k$ converging to zero and $\alpha_k$ bounded.

From this analysis, we conclude that method 3 is the best choice for implementation.
Prior to providing implementation details, however, we briefly consider a simple multistage 
extension that can capture the exact first and second moments of position and velocity
simultaneously at the expense of introducing a numerical correlation between them.

\subsection{Multistage Splitting}
\label{multistage_splitting}

Inspired by the work of Leimkuhler and Matthews,\cite{leimkuhler13} we consider a generalization
of the splitting method considered in Section~\ref{numerical_integration_of_the_GLE} 
in which the position and velocity updates are further split,
\begin{eqnarray*}
 V^{n+1/4}_i &=& V^{n}_i + \frac{\Delta t}{2 m_i} F^c_i\left(X^n\right) + (1 - \xi) \frac{\Delta t}{2 m_i} \sum \limits_{k=1}^{N_k} S^n_{i,k} \\
 X^{n+1/2}_i &=& X^{n}_i + \frac{\Delta t}{2} V^{n+1/4}_i \\
 V^{n+1/2}_i &=& V^{n+1/4}_i + \xi \frac{\Delta t}{2 m_i} \sum \limits_{k=1}^{N_k} S^n_{i,k} \\
 S^{n+1}_{i,k} &=& \theta_{k} S^{n}_{i,k} - \left(1-\theta_{k}\right) c_k V^{n+1/2}_i + \alpha_{k} \sqrt{2 \text{k}_B\text{T} c_k } B^n_{i,k} \\
 V^{n+3/4}_i &=& V^{n+1/2}_i + \xi \frac{\Delta t}{2 m_i} \sum \limits_{k=1}^{N_k} S^{n+1}_{i,k} \\
 X^{n+1}_i &=& X^{n+1/2}_i + \frac{\Delta t}{2} V^{n+3/4}_i \\
 V^{n+1}_i &=& V^{n+3/4}_i + \frac{\Delta t}{2 m_i} F^c_i\left(X^{n+1}\right) + (1 - \xi) \frac{\Delta t}{2 m_i} \sum \limits_{k=1}^{N_k} S^{n+1}_{i,k} 
\end{eqnarray*}
In the special case that $\xi = 0$, we have $V_i^{n+1/4} = V_i^{n+1/2} = V_i^{n+3/4}$, the two updates of $X$ can be combined, and we recover our original splitting method.

Repeating the analysis in Section~\ref{error_analysis}
for the harmonic oscillator with a single memory term,
we find
\[
\langle (V^{n})^2 \rangle = \frac{\text{k}_B\text{T} }{m} \frac{\Delta t \alpha^2(4 - (\Delta t \omega)^2 \xi)}{4 (1 - \theta)^2}, \quad 
\langle X^n V^n \rangle = 0, \quad
\langle V^n S^n \rangle = 0, 
\]
\[
\langle (X^{n})^2 \rangle =  \frac{\text{k}_B\text{T} }{m \omega^2} \frac{\Delta t \alpha^2 ( c \Delta t (1 - \theta)^2 (4 - (\omega \Delta t)^2 \xi^2) - 2 m (1 - \theta^2)(4 - (\omega \Delta t)^2 \xi))}{(1 - \theta)^2 (c \Delta t (1 - \theta)^2 (4 - (\omega \Delta t)^2 \xi ) - 2 m (1 - \theta^2)(4 - (\omega \Delta t)^2))}, \]
\[
\langle X^n S^n \rangle = \frac{4 \Delta t^2 \alpha^2 (1 - \xi) c \, \text{k}_B\text{T} }
{c \Delta t (1 - \theta)^2 (4 - (\omega \Delta t)^2 \xi)  - 2 m (1 - \theta^2) (4 - (\omega \Delta t)^2)},
\]
\[ \text{and} \quad
\langle (S^{n})^2 \rangle = \frac{c \, \text{k}_B\text{T}}{\tau} \frac{ 4 \tau \alpha^2 m (4 - (\omega \Delta t)^2)}{2 m (1 - \theta^2)(4 - (\omega \Delta t)^2) -  c \Delta t (1 - \theta)^2 (4 - (\omega \Delta t)^2 \xi) }.
\]
In the special case that $\xi = 1$, we can simplify these expressions to 
obtain
\[
\langle (V^{n})^2 \rangle = \frac{\text{k}_B\text{T} }{m} \frac{\Delta t \alpha^2(4 - (\Delta t \omega)^2)}{4 (1 - \theta)^2}, \qquad
\langle (X^{n})^2 \rangle =  \frac{\text{k}_B\text{T} }{m \omega^2} \frac{\Delta t \alpha^2}{(1 - \theta)^2}, \quad
\langle X^n V^n \rangle = 0,
\]
\[ 
\langle X^n S^n \rangle = 0, \quad
\langle V^n S^n \rangle = 0, \quad
\text{and} \quad
\langle (S^{n})^2 \rangle = \frac{c \, \text{k}_B\text{T}}{\tau} \frac{ 4 \tau \alpha^2 m}{2 m (1 - \theta^2) -  c \Delta t (1 - \theta)^2 }.
\]
From this analysis, we conclude that we obtain the correct second moment for
$X$ for any method with $\xi = 1$ and $\Delta t \alpha^2 = (1 - \theta)^2$.  
To guarantee the correct second moment for $V$, the choice of $\theta$ and 
$\alpha$ becomes $\omega$ dependent.  As was discussed in the previous section, 
the parameters prescribed by methods 1 and 3 using the original splitting 
have a similar behavior but with the roles of $X$ and $V$ reversed.

It is then tempting to formulate a method that exactly preserves the second 
moments of both $X$ and $V$ at the same time.  It turns out that this is 
possible by simply shifting where we observe $V$, using either 
$V_i^{n+1/4}$ or $V_i^{n+3/4}$ in the multistage splitting method above.
For example, consider the following asymmetric method, with $\xi = 1$,
\begin{eqnarray*}
 X^{n+1/2}_i &=& X^{n}_i + \frac{\Delta t}{2} V^{n}_i \\
 V^{n+1/2}_i &=& V^{n}_i + \frac{\Delta t}{2 m_i} \sum \limits_{k=1}^{N_k} S^n_{i,k} \\
 S^{n+1}_{i,k} &=& \theta_{k} S^{n}_{i,k} - \left(1-\theta_{k}\right) c_k V^{n+1/2}_i + \alpha_{k} \sqrt{2 \text{k}_B\text{T} c_k } B^n_{i,k} \\
 V^{n+3/4}_i &=& V^{n+1/2}_i + \frac{\Delta t}{2 m_i} \sum \limits_{k=1}^{N_k} S^{n+1}_{i,k} \\
 X^{n+1}_i &=& X^{n+1/2}_i + \frac{\Delta t}{2} V^{n+3/4}_i \\
 V^{n+1}_i &=& V^{n+3/4}_i + \frac{\Delta t}{m_i} F^c_i\left(X^{n+1}\right) \\
\end{eqnarray*}
For this method we obtain,
\[
\langle (V^{n})^2 \rangle = \frac{\text{k}_B\text{T} }{m} \frac{\Delta t \alpha^2}{(1 - \theta)^2}, \qquad
\langle (X^{n})^2 \rangle =  \frac{\text{k}_B\text{T} }{m \omega^2} \frac{\Delta t \alpha^2}{(1 - \theta)^2}, \quad
\langle X^{n} V^{n} \rangle = \frac{- \Delta t^2 \alpha^2 \text{k}_B\text{T}}{2 m (1- \theta)^2},
\]
\[
\langle X^n S^n \rangle = 0, \quad
\langle V^n S^n \rangle = 0, \quad
 \text{and} \quad
\langle (S^{n})^2 \rangle = \frac{c \, \text{k}_B\text{T}}{\tau} \frac{ 4 \tau \alpha^2 m}{2 m (1 - \theta^2) -  c \Delta t (1 - \theta)^2 }.
\]
If we use a method with $\Delta t \alpha^2 = (1 - \theta)^2$, we find that
we obtain the exact moments for $X$ and $V$, but we have introduced an 
$\mathcal{O}(\Delta t)$ correlation between $X$ and $V$.  This is in 
contrast to the symmetric methods where this correlation is identically zero.

\section{Implementation Details}\label{implementation_details}

Method 3, as detailed in Section \ref{error_analysis} 
has been implemented in the LAMMPS software package.  It can be applied in conjunction 
with all conservative force fields supported by LAMMPS.  There are a number of details 
of our implementation worth remarking on concerning random number generation,
initial conditions on the extended variables, and the conservation of total
linear momentum.  

The numerical integration scheme requires the generation of Gaussian random 
numbers, by way of $B^n_{i,k}$ in Equation \ref{s_update}. By default, all 
random numbers are drawn from a uniform distribution with the same mean and 
variance as the formally required Gaussian distribution.  This distribution 
is chosen to avoid the generation of numbers that are arbitrarily large, or 
more accurately, arbitrarily close to the floating point limit.  The 
generation of such large numbers may lead to rare motions that result in the 
loss of atoms from a periodic simulation box, even at low temperatures.  Atom 
loss occurs if, within a single time step, the change in one or more of an 
atom's position coordinates is updated to a value that results in it being 
placed outside of the simulation box after periodic boundary conditions are 
applied. A uniform distribution can be used to guarantee that this will not 
happen for a given temperature and time step.  However, for the sake of 
mathematical rigor, the option remains at compile-time to enable the use of 
the proper Gaussian distribution with the caveat that such spurious motions 
may occur.  Should the use of this random number generator produce a 
trajectory in which atom loss occurs, a simple practical correction may be to 
use a different seed and/or a different time step in a subsequent simulation. 
It is worth noting that the choice of a uniform random number distribution has 
been rigorously justified by D\"unweg and Paul\cite{dunweg91} for a number of 
canonical random processes, including one described by a conventional Langevin 
equation.  We anticipate that a similar result may hold for the extended 
variable GLE presented in this manuscript.

With respect to the initialization of the extended variables, it is frequently 
the case in MD that initial conditions are drawn from the equilibrium 
distribution at some initial temperature.  Details of the equilibrium 
distribution for the extended system are presented in Section 2 of an article 
by Ottobre and Pavliotis.\cite{ottobre11}  In our implementation, we provide 
the option to initialize the extended variables either based upon this 
distribution, or with zero initial conditions (i.e., the extended system at 
zero temperature).  As it is typically more relevant for MD simulations, the 
former is enabled by default and used in the generation of the results in this 
paper.  

Conservation of the total linear momentum of a system is frequently a 
desirable feature for MD trajectories.  For deterministic forces, this can be 
guaranteed to high precision through the subtraction of the velocity of the 
center of mass from all particles at a single time step. In the presence of 
random forces such as those arising in GLD, a similar adjustment must be made 
at each time step to prevent the center of mass from undergoing a random 
acceleration.  While it is not enabled by default, our implementation provides 
such a mechanism that can be activated.  When active, the average of the 
forces acting on all extended variables is subtracted from each individual 
extended variable at each time step.  While this is a computationally 
inexpensive adjustment, it may not be essential for all simulations.  

\section{Results}\label{results}

Throughout this section, results will be presented, primarily in terms of the 
integrated velocity autocorrelation function (VAF).  This quantity is 
calculated using ``block averaging'' and ``subsampling'' of the integrated 
trajectories for computational convenience.\cite{allenbook}  Error bars are 
derived from the standard deviation associated with a set of independently 
generated trajectories.

\begin{figure}[ht]
 \centering
 \graphicsmacro{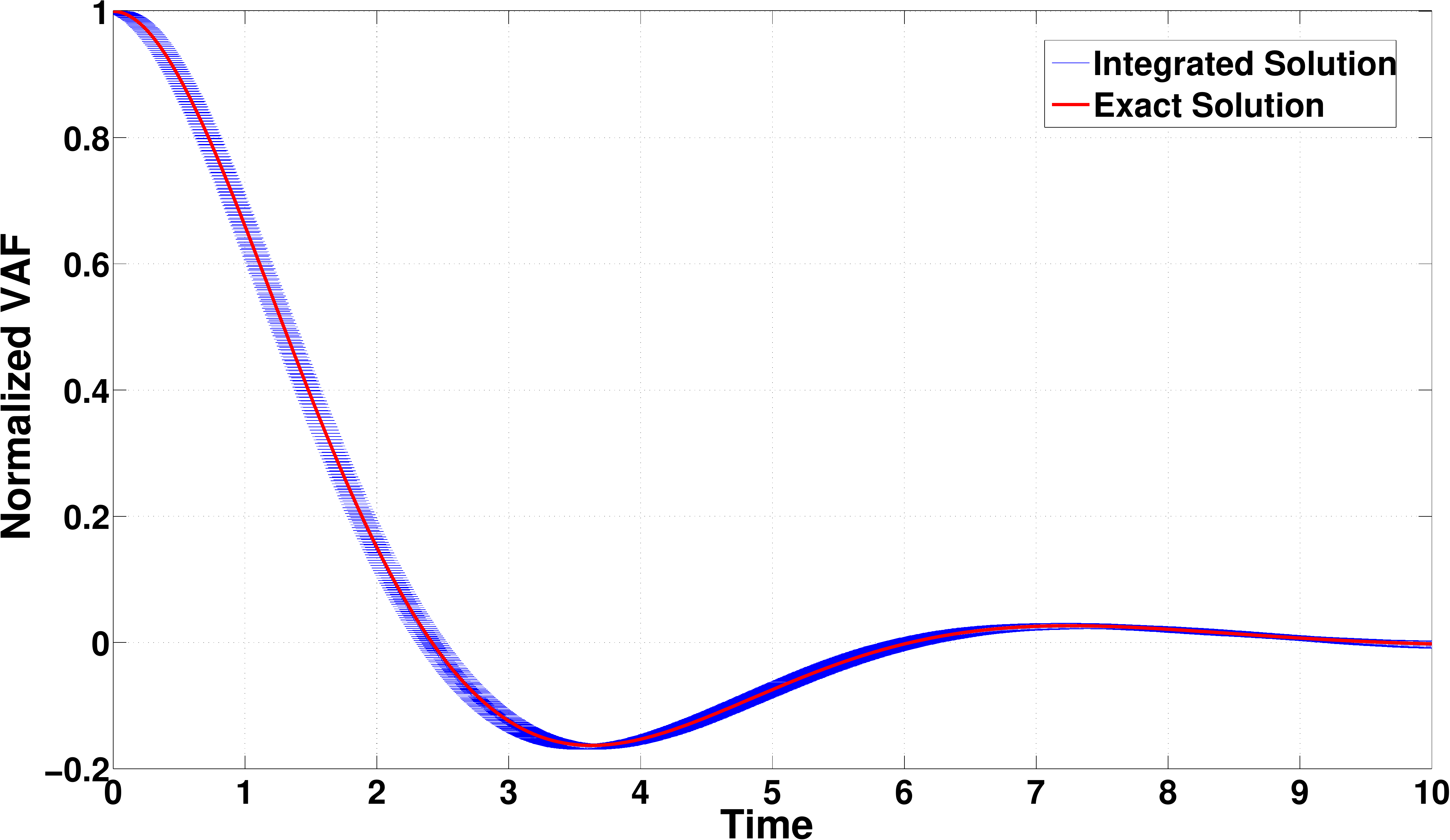} 
 \graphicsmacro{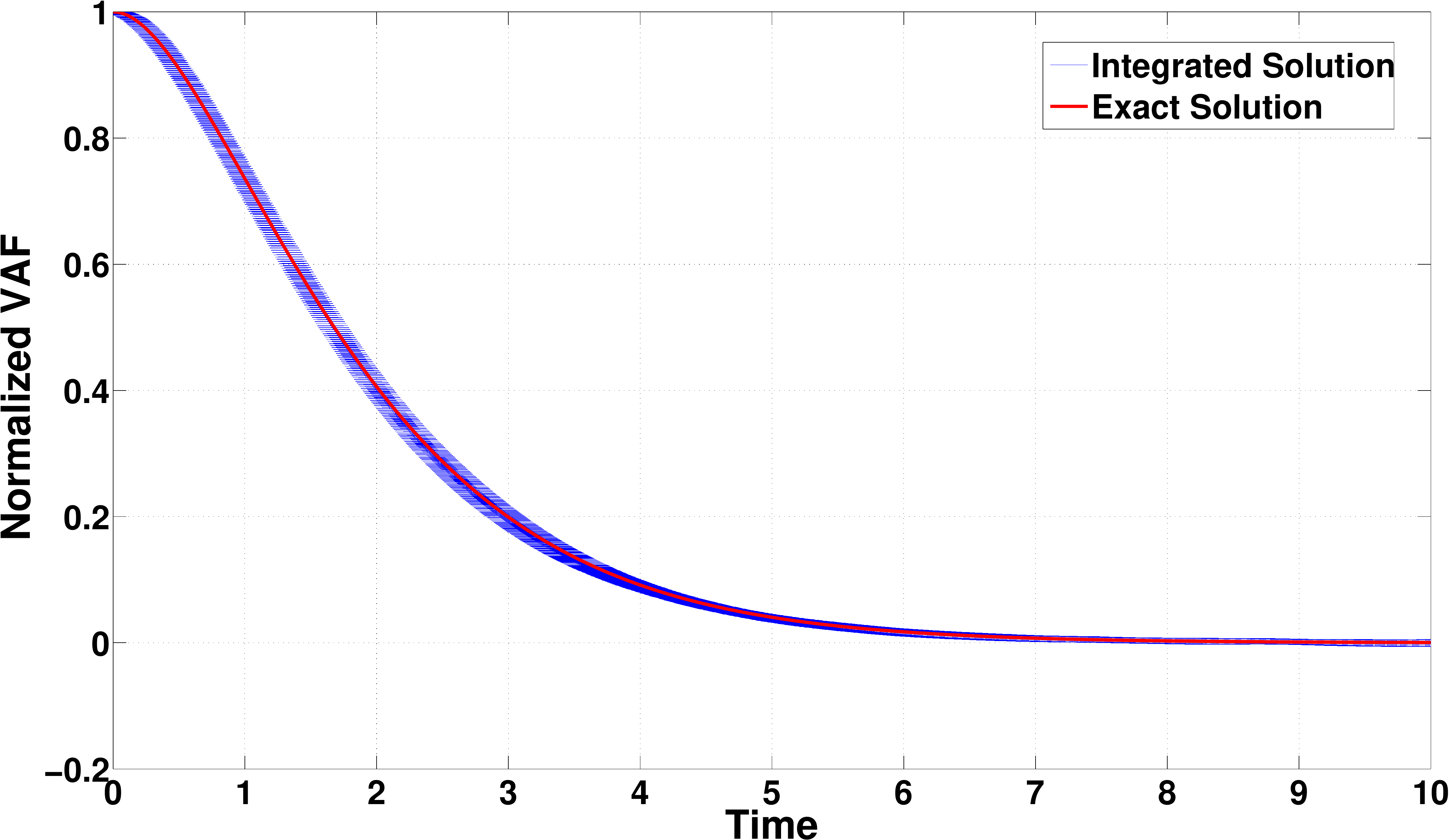} 
 \graphicsmacro{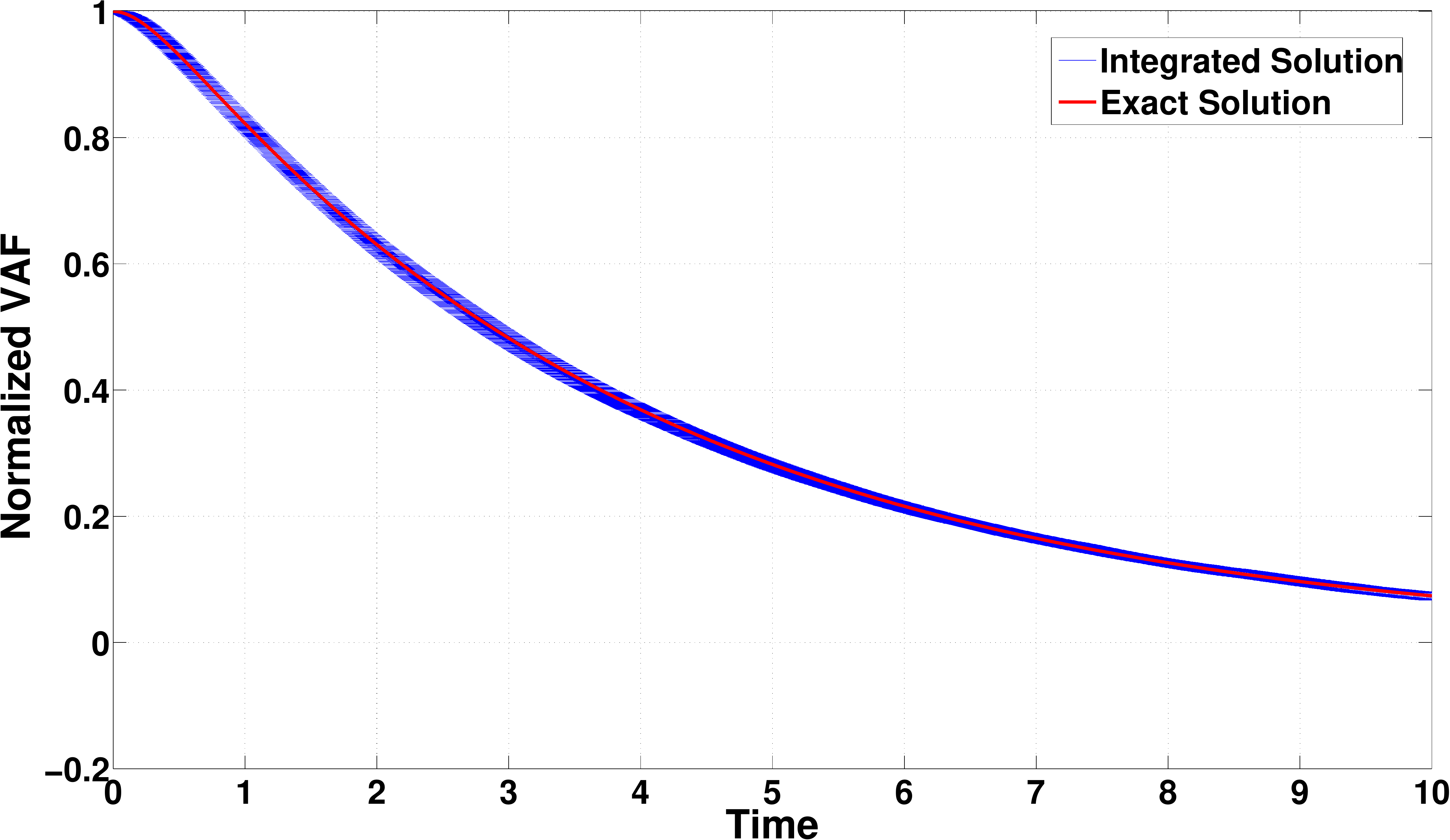} 
 \caption{Normalized VAF for a single Prony series mode in the i.) underdamped ($c=1$, $\tau=1$), ii.) critically damped ($c=0.5$, $\tau=0.5$), and iii.) overdamped limits ($c=0.25$, $\tau=0.25$).  
A time step of $\Delta t=0.01$ is used for all runs, and error bars are drawn based upon a sample of 10,000 walkers over 10 independent runs. \label{single_mode_fig}}
\end{figure}

We begin by presenting results that validate our time integration scheme for a 
GLE in the absence of a conservative force with a single mode Prony series 
kernel.  In this case, the GLE is analytically soluble.\cite{berne66}  We 
consider the normalized VAF as a metric for comparison.  For a kernel of the 
form
\begin{equation}
 \label{single_mode_kernel}
 \Gamma(t) = \frac{c}{\tau} \exp\left[\frac{-t}{\tau}\right],~~~t \geq 0,
\end{equation}
the normalized VAF takes the following form:
\begin{equation}
 \label{single_mode_vaf}
 \frac{\langle V(t) V(0) \rangle}{\langle V(0) V(0) \rangle} = \begin{cases}
\exp\left[\frac{-t}{2\tau}\right]\left( \cos(\Omega t) + \frac{1}{2\tau\Omega} \sin(\Omega t)\right) & \text{for $\Omega\neq0$} \\
\exp\left[\frac{-t}{2\tau}\right]\left( 1 + \frac{t}{2\tau} \right) & \text{for $\Omega=0$}
\end{cases},~~~\Omega = \sqrt{c/\tau - 1/4\tau^2}.
\end{equation}
Making an analogy with the canonical damped harmonic oscillator, we consider 
three scenarios, i. underdamped (real $\Omega$), 
ii. critically damped ($\Omega=0$), and iii. overdamped (imaginary $\Omega$).
In Figure \ref{single_mode_fig}, we demonstrate that we can recover all three 
regimes using our integrator.

\begin{figure}[ht]
 \centering
 \graphicsmacro{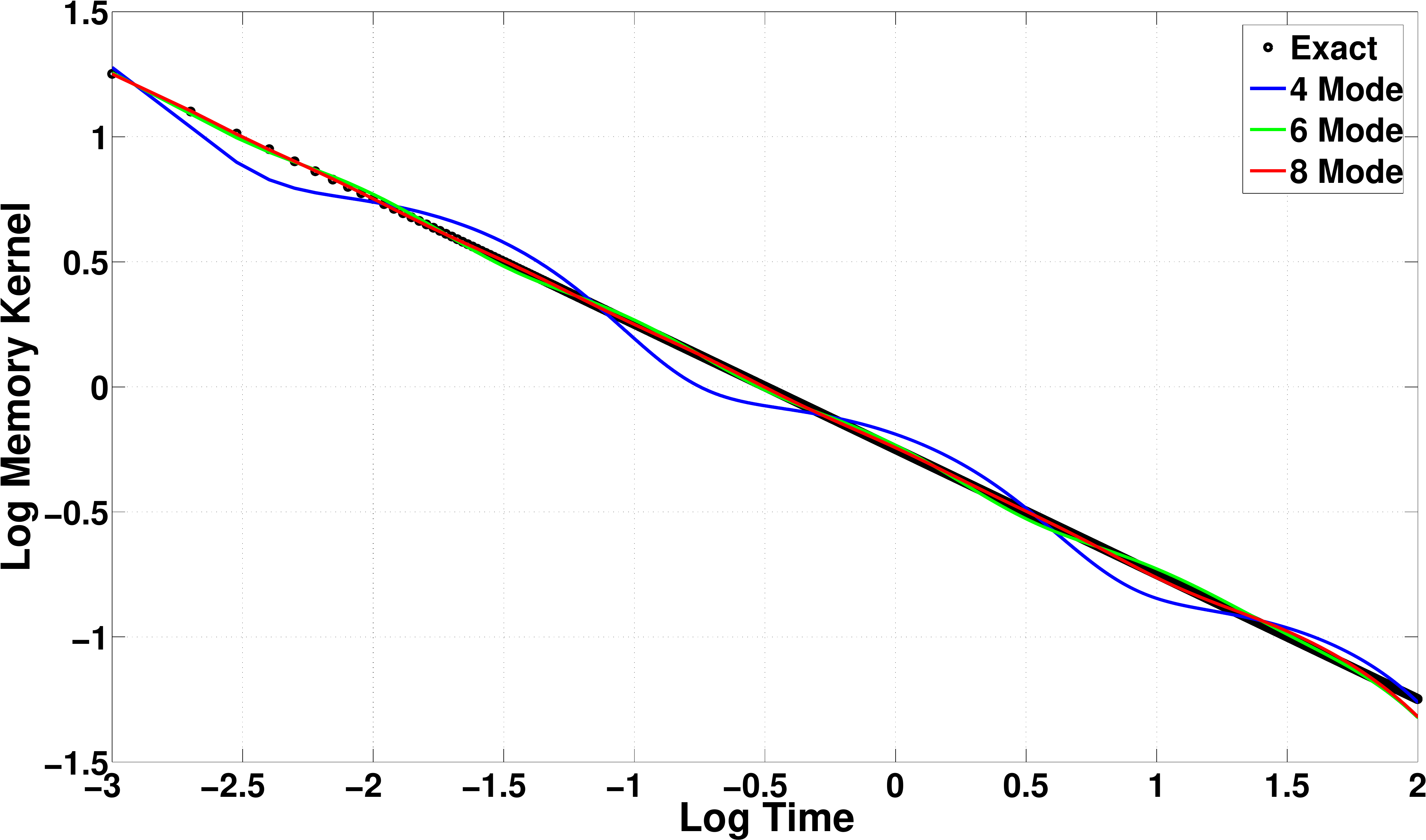}
 \caption{A Prony series fit of the power law memory kernel in Equation \ref{harmonic_gle} for $\gamma_\lambda=1$, $\lambda=0.5$ with an 
increasing number of modes.\label{kernel_fit_fig}}
\end{figure}

To validate our integration scheme in the presence of a conservative force, we 
next consider GLD with an external potential. The analytic solution of the GLE 
with a power law memory kernel and a harmonic confining potential has been 
derived by other authors.\cite{desposito09}  In the cited work, the GLE is 
solved in the Laplace domain, yielding correlation functions given in terms of 
a series of Mittag-Leffler functions and their derivatives.  Here, we apply 
our integration scheme to a Prony series fit of the power law kernel and 
demonstrate that our results are in good agreement with the exact result over 
a finite time interval.  The GLE that we intend to model has the form:
\begin{equation}
 \label{harmonic_gle}
 dV(t) = -\omega_0^2 X(t)dt -\int \limits_{0}^{t} \frac{\gamma_\lambda}{\Gamma(1-\lambda)} (t-s)^{-\lambda} V(s)ds dt + M^{-1}F^r(t)dt,~~~0 < \lambda < 1
\end{equation}
We begin by constructing a Prony series representation of the memory kernel.  
As it exhibits a power law decay in the Laplace domain, as well, a Prony series
representation will have strong contributions from modes decaying over a 
continuum of time scales.   To this end, rather than relying on a non-linear 
fitting procedure to choose values for $\tau_k$, we assume logarithmically 
spaced values from $\Delta t/10$ to $10 N_t \Delta t$.  By assuming the form 
of each exponential, the Prony series fit reduces to a simple linear least 
squares problem, that we solve using uniformly spaced data over an interval 
that is two decades longer than the actual simulation.  In 
Figure \ref{kernel_fit_fig}, the Prony series fit of the memory kernel for 
$\lambda=0.5$ is compared to its exact value.  

\begin{figure}[ht]
 \centering 
 \graphicsmacro{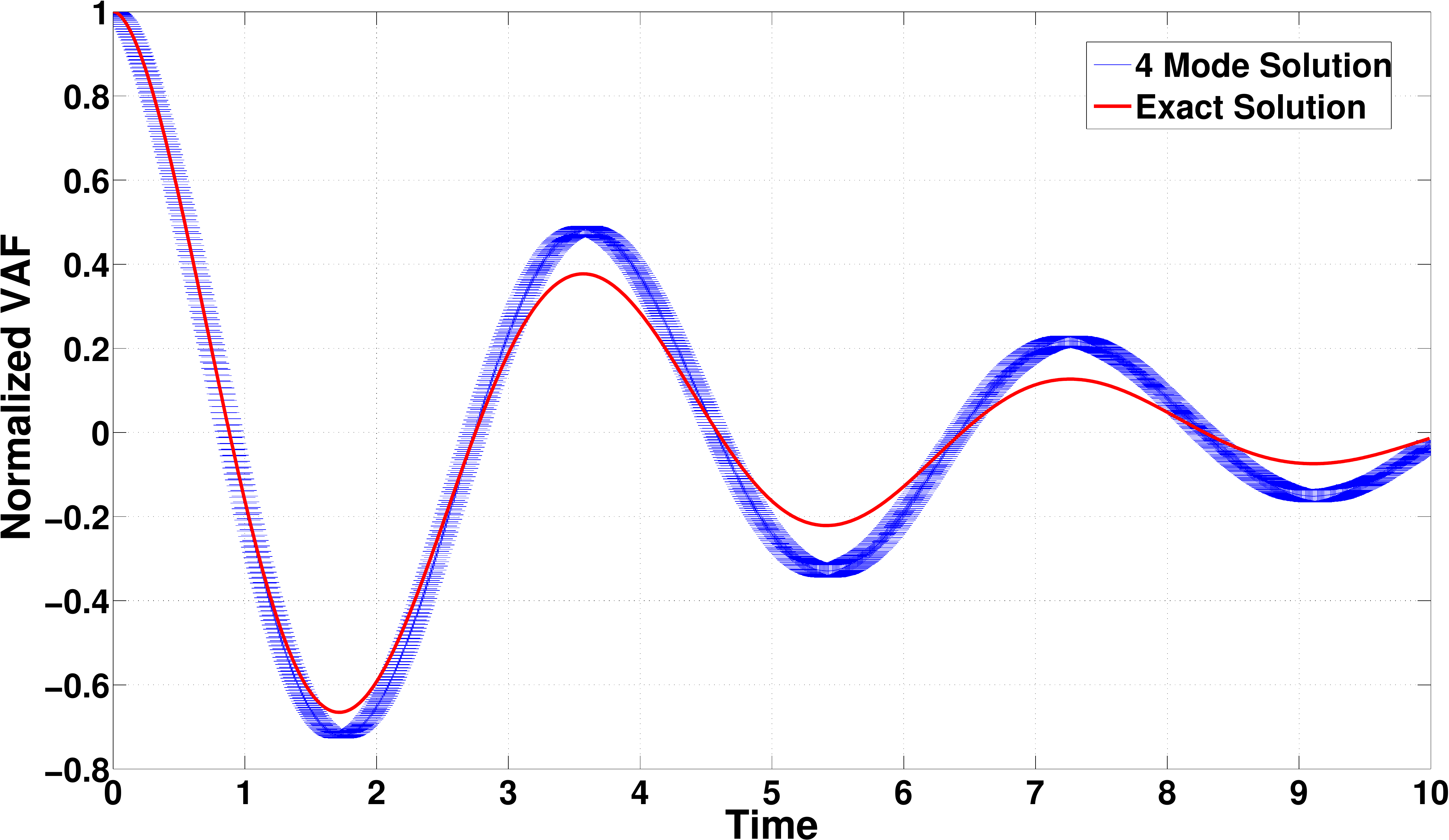} \graphicsmacro{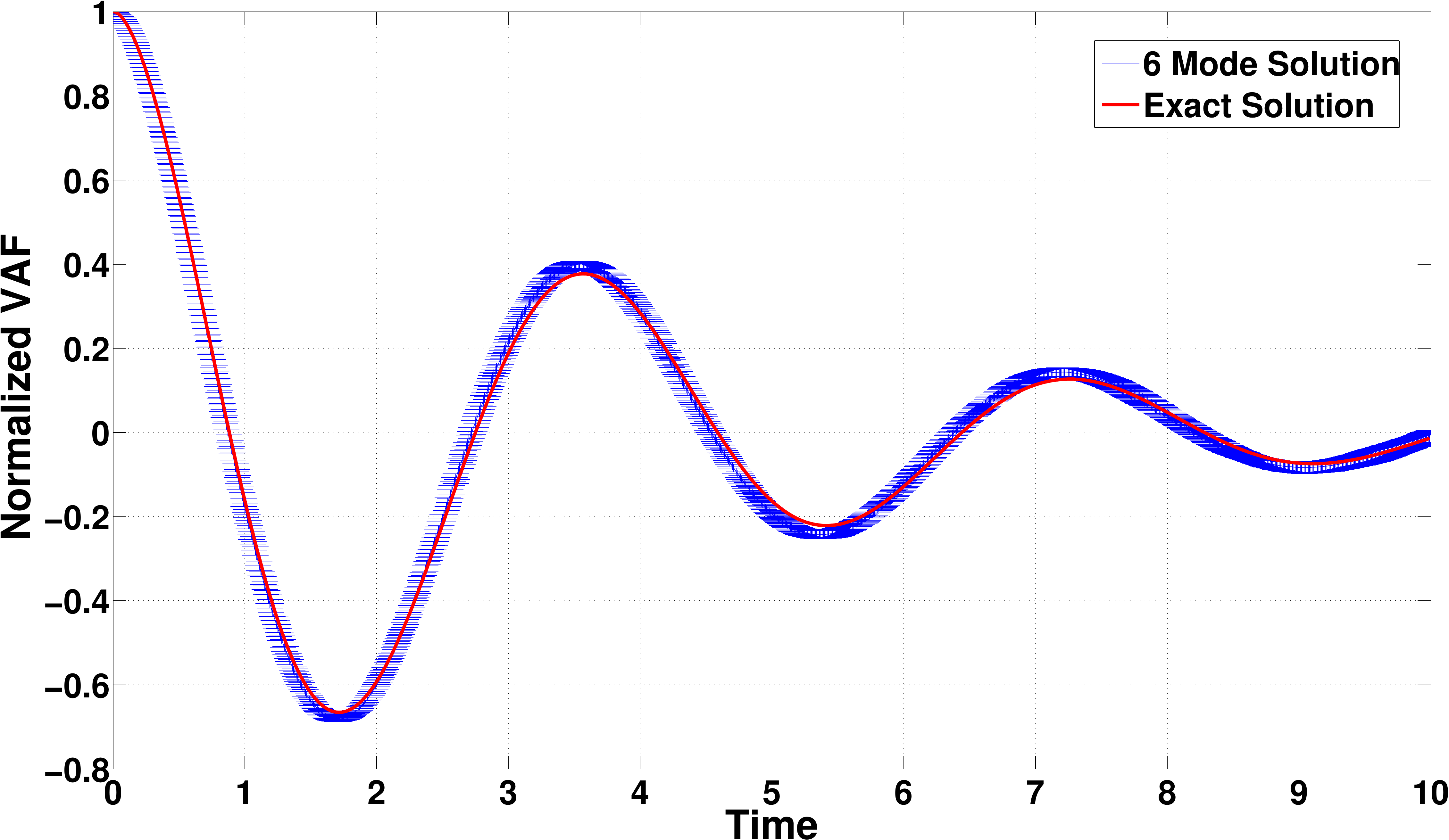} \\
 \graphicsmacro{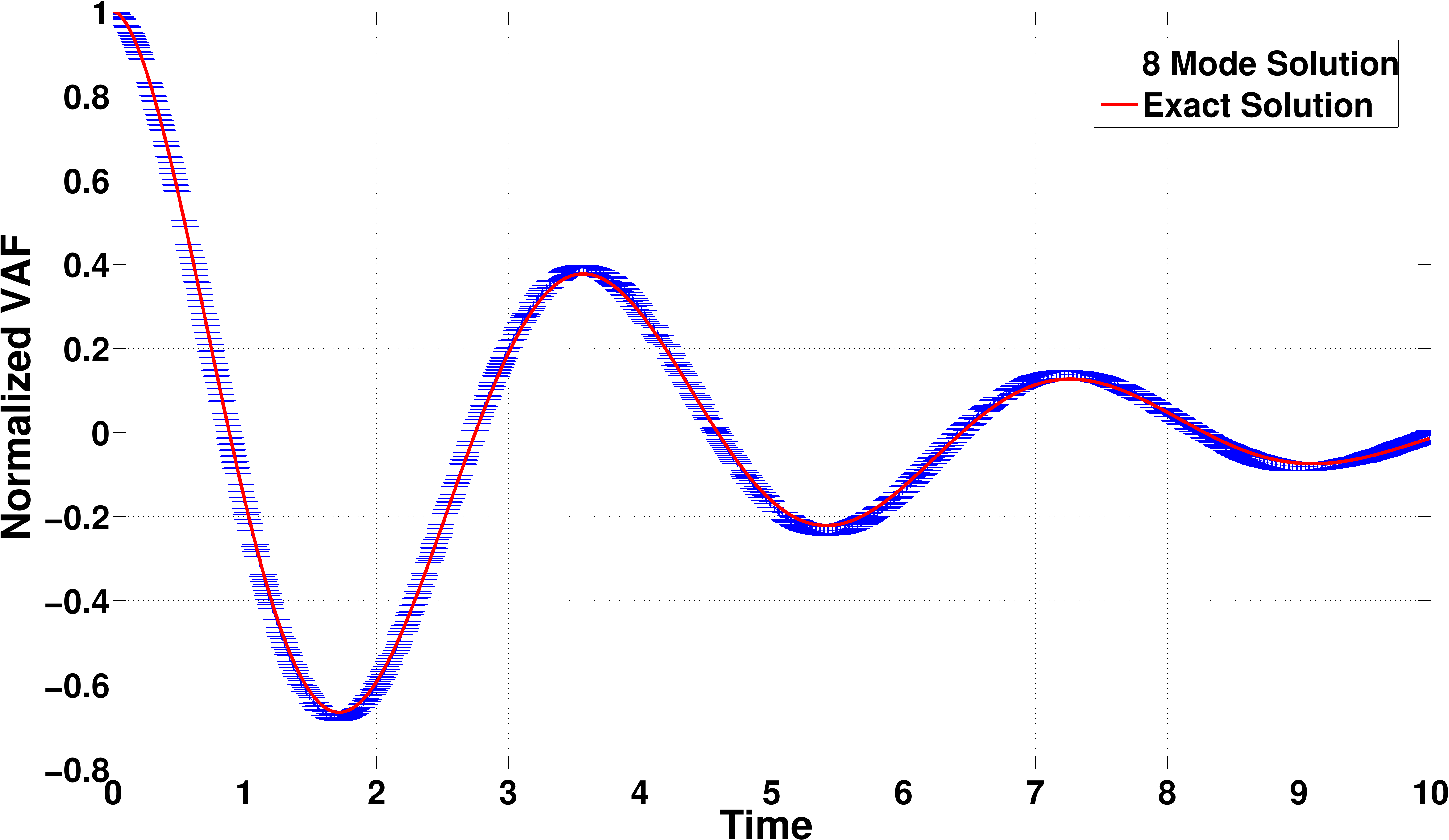}
 \caption{Comparison of numerically integrated results to the exact solution of Equation \ref{harmonic_gle} for $\gamma_\lambda=1$, $\lambda=0.5$, and $\omega_0=1.4$. 
A time step of $\Delta t=0.01$ is used, and error bars are drawn based upon a sample of 10,000 walkers over 10 independent runs with $\Delta t=0.01$.\label{harmonic_soln_fig}}
\end{figure}

\begin{figure}[ht]
 \centering 
 \graphicsmacro{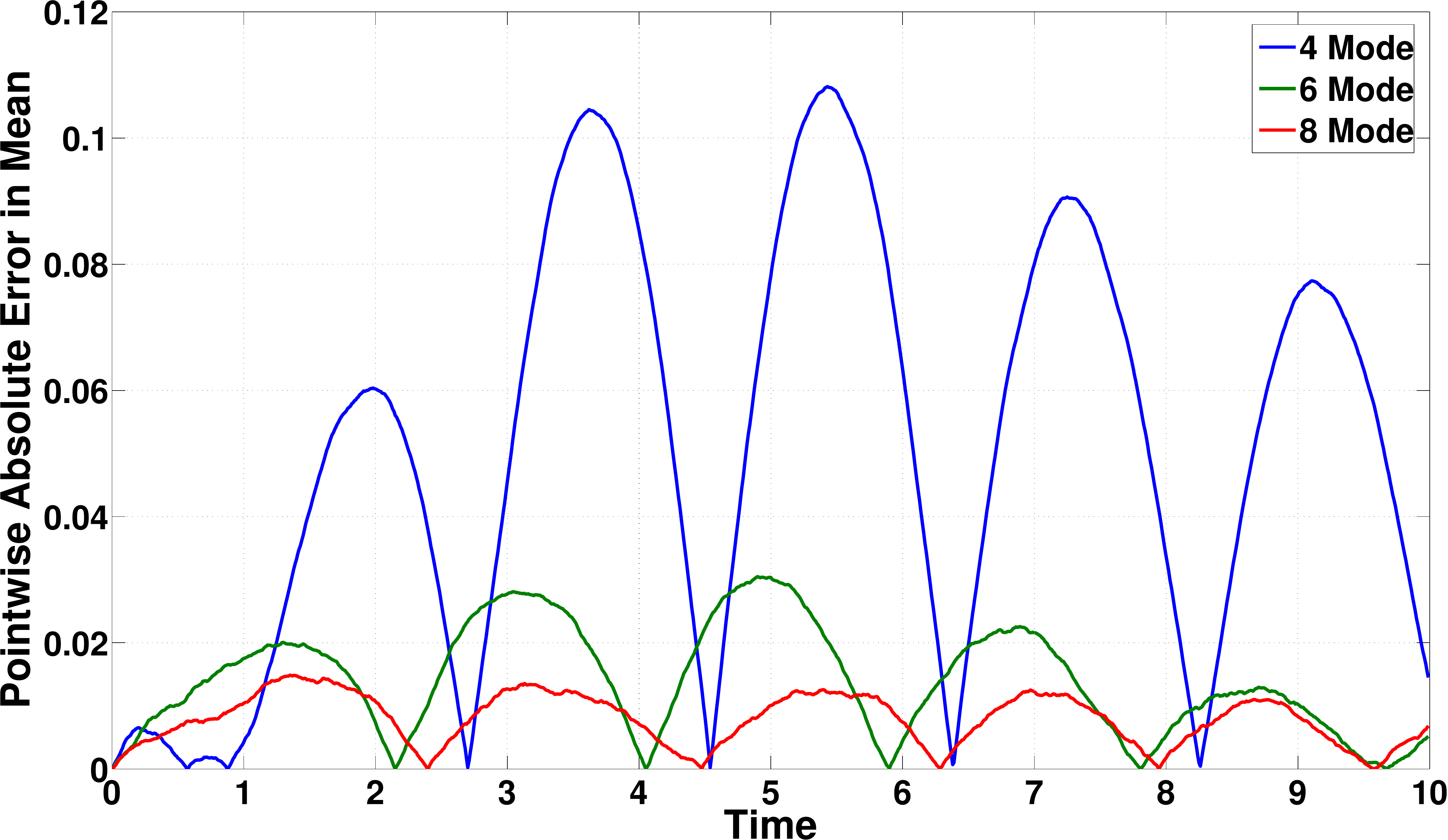}
 \caption{Pointwise absolute error in the normalized velocity autocorrelation function for the same conditions as Figure \ref{harmonic_soln_fig}. 
 Error is computed with respect to the mean of VAFs computed from 10 independent runs.\label{harmonic_error_fig}}
\end{figure}

Here, the maximum relative error is $\sim 10\%$, while it is $\sim 1\%$ for 
$N_k=8$.  In Figure \ref{harmonic_soln_fig}, the normalized VAF computed via 
numerical integration  of the extended GLE with a variable number of modes is 
shown compared to the exact result for some of the parameters utilized in an 
article by Desp{\'o}sito and Vi{\~n}ales.\cite{desposito09}  It seems evident 
that the accuracy of the integrated velocity distribution improves relative to 
the exact velocity distribution as the number of terms in the Prony series fit 
increases. This is quantified in Figure \ref{harmonic_error_fig}, in which the 
pointwise absolute error in the integrated VAF is illustrated.  

\begin{figure}[ht]
 \centering
 \graphicsmacro{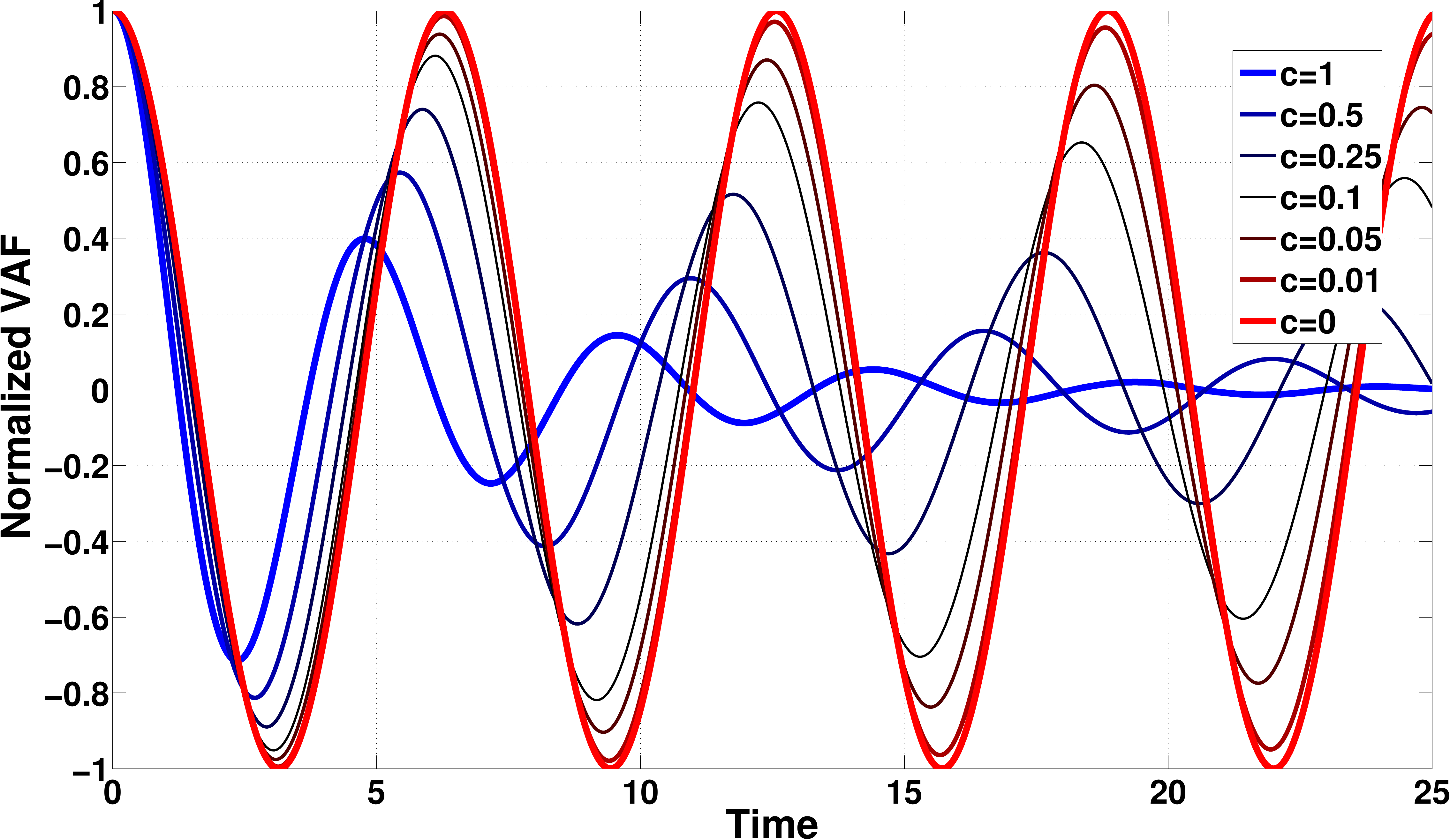}
 \caption{Illustration of the smooth/stable manner in which the proposed integrator approaches the $c=0$ (Newtonian) limit from $c=1$ for a single Prony mode with $\tau=1$. The particle is 
harmonically confined with $\omega_0=1$, and the expected period of oscillation is restored for $c=0$. \label{zero_coeff_fig}}  
\end{figure}

Next, we demonstrate that our implementation is robust in certain limits of 
the GLE - particularly the Langevin and zero coefficient limits. As our 
implementation is available in a public domain code with a large user base, 
developing a numerical method that is robust to a wide array of inputs is 
essential. For the zero coefficient limit we consider a harmonically confined 
particle experiencing a single mode Prony series memory kernel of the 
following form:
\begin{equation}
 \label{single_mode_harmonic_gle}
 dV(t) = -\omega_0^2 X(t)dt -\int \limits_{0}^{t} \frac{c}{\tau} e^{-(t-s)/\tau} V(s)ds dt + M^{-1}F^r(t)dt,~~~0 < \lambda < 1
\end{equation}
The initial conditions on $X$ and $V$ are drawn from a thermal distribution at 
$T=1$.  In the limit that $c\to0$, we expect the integrated normalized VAF to 
approach that of a set of deterministic harmonic oscillators.  The initial 
conditions on $X$ and $V$ are drawn from a thermal distribution at $T=1$, 
giving the integrated result some variance, even in the 
Newtonian/deterministic limit.  In Figure \ref{zero_coeff_fig}, that this 
limit is smoothly and stably approached is illustrated.  

\begin{figure}[ht]
 \centering
 \graphicsmacro{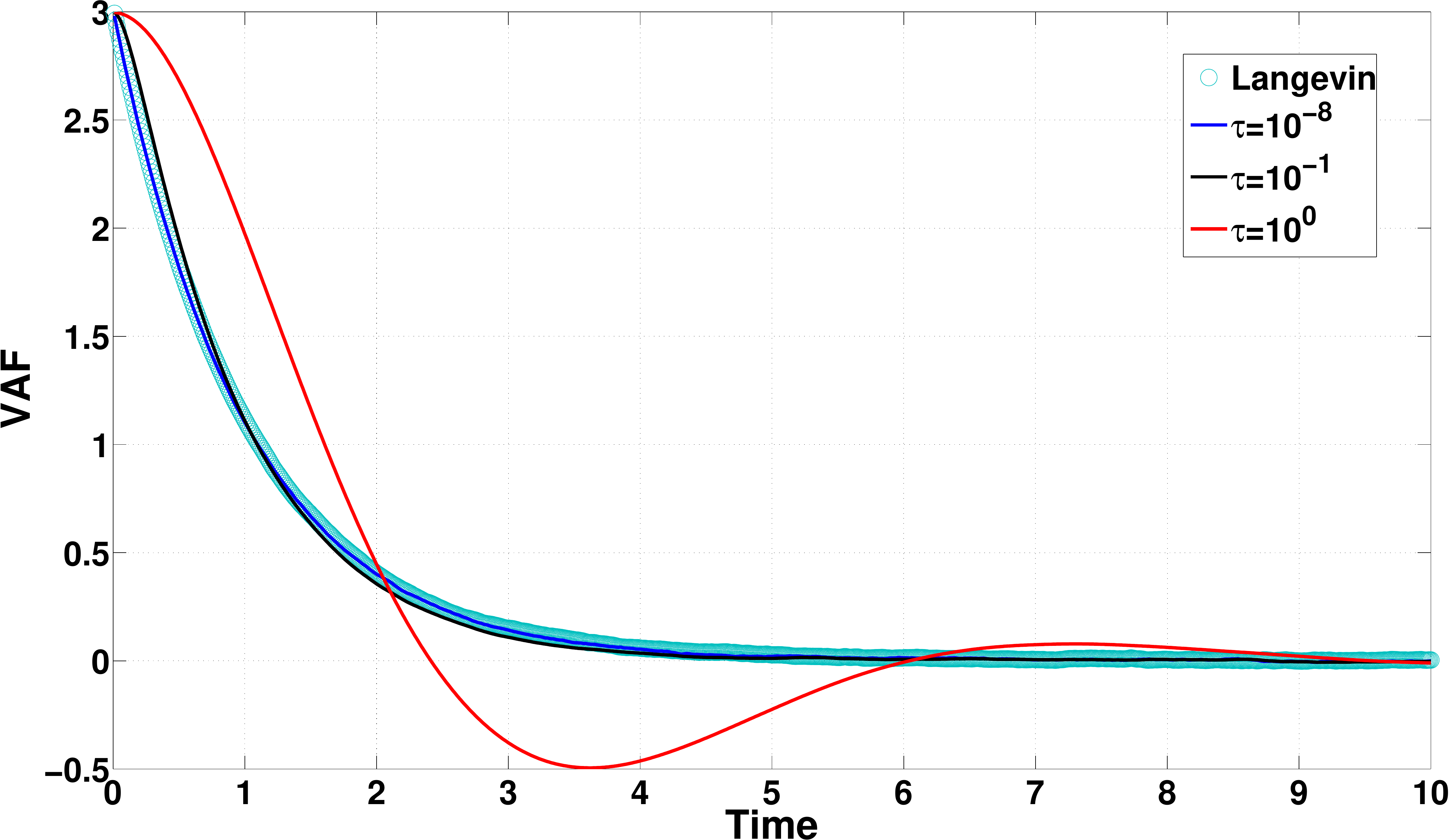}
 \caption{Illustration of the smooth/stable manner in which the proposed integrator approaches the $\tau=0$ (Langevin) limit from $\tau=1$ for a single Prony mode.  The particle is subject to 
no conservative forces, so the resultant dynamics correspond to an Ornstein-Uhlenbeck process.  The `exact' Langevin limit was integrated using {\it fix langevin} in LAMMPS. \label{langevin_lim_fig}}
\end{figure}

In the exact $c=0$ limit, oscillations in the VAF occur at the natural 
frequency of the confining potential, $\omega_0=1$, whereas for non-zero $c$, 
these oscillations are damped in proportion to $c$, as one may expect on the 
basis of intuition.  The Langevin limit is illustrated next.  To this end, we 
utilize the same single mode Prony series memory kernel, but remove the 
confining potential (i.e., $\omega_0=0$).  We have done so to ensure that the 
Langevin limit yields an Ornstein-Uhlenbeck process.  In 
Figure \ref{langevin_lim_fig}, we illustrate that this limit is also smoothly 
and stably approached.  

The result for the Langevin limit itself was integrated using the existing 
{\it fix langevin} command in LAMMPS.  For $\tau=10^0$, the GLE yields results 
that differ from the Langevin limit as one might expect on the basis of the 
previous results.  However, even for $\tau=10^{-1}$, the GLE yields results 
that are close to that of the Langevin limit, as the resultant numerical 
method retains some GLE-like behavior.  However, for $\tau=10^{-8}$ and 
$\Delta t=0.01$, the parameter $\theta=\exp\left[-10^6\right]$, is below 
machine epsilon.  As mentioned in Section \ref{error_analysis}, this results 
in a method that is indistinguishable from conventional Langevin dynamics, and 
consequently, the integrated dynamics are indistinguishable.  This result 
demonstrates that the method is robust, even if the user `accidentally' 
crosses into a Langevin-like regime.  We again emphasize that methods 1 and 2,
considered earlier, will not stably approach this limit.

To demonstrate capability in reduced order modeling, we next invert the VAF 
associated with a set of trajectories generated in the presence of a 
conservative force, to construct a memory kernel that will reproduce the same 
dynamics without such a force.  In other words, we seek an effective memory kernel, 
$\Gamma_{eff}(t)$, such that a simplified GLE of the form:
\begin{equation}
 \label{force_free_gle}
 dV(t) = -\int \limits_{0}^{t} \Gamma_{eff}(t-s) V(s)ds dt + M^{-1}F^r(t)dt
\end{equation}
reproduces the VAF of a GLE of the more general form given in Equation \ref{cont_gle}.

The starting point for this procedure is the observation that for Equation \ref{force_free_gle},
the VAF and memory kernel have the following relationship in the Laplace domain:\cite{mason97}
\begin{equation}
 \langle \widetilde{V}(\sigma) V(0) \rangle = \frac{\text{k}_B\text{T}}{m\sigma + \widetilde{\Gamma}_{eff}(\sigma)}.
\end{equation}
Here, tildes indicate that the associated time domain quantities have been 
Laplace transformed, with transform domain variable $\sigma$.  Given $\langle \widetilde{V}(\sigma) V(0) \rangle$ 
for some process whose VAF we wish to reproduce via integration of Equation \ref{force_free_gle}, 
$\widetilde{\Gamma}_{eff}(\sigma)$ can be resolved algebraically, and its time domain 
form can be found via an inverse Laplace transform:
\begin{equation}
 \label{kernel_fit_eqn}
 \Gamma_{eff}(t) = \mathcal{L}^{-1} \left\{ \frac{\text{k}_B\text{T}}{\langle \widetilde{V}(\sigma) V(0) \rangle} - m\sigma\right\}(t)
\end{equation}
This effective memory kernel can then be fit to a Prony series, and Equation \ref{force_free_gle}
can be integrated using the numerical method developed in this manuscript.  In Figure 
\ref{harmonic_kernel_fit}, we demonstrate that we can use this inversion procedure to 
take the exact VAF generated by Equation \ref{harmonic_gle}, and compute a $\Gamma_{eff}(t)$ 
for Equation \ref{force_free_gle} that reproduces it:
\begin{figure}[ht]
 \centering
 \graphicsmacro{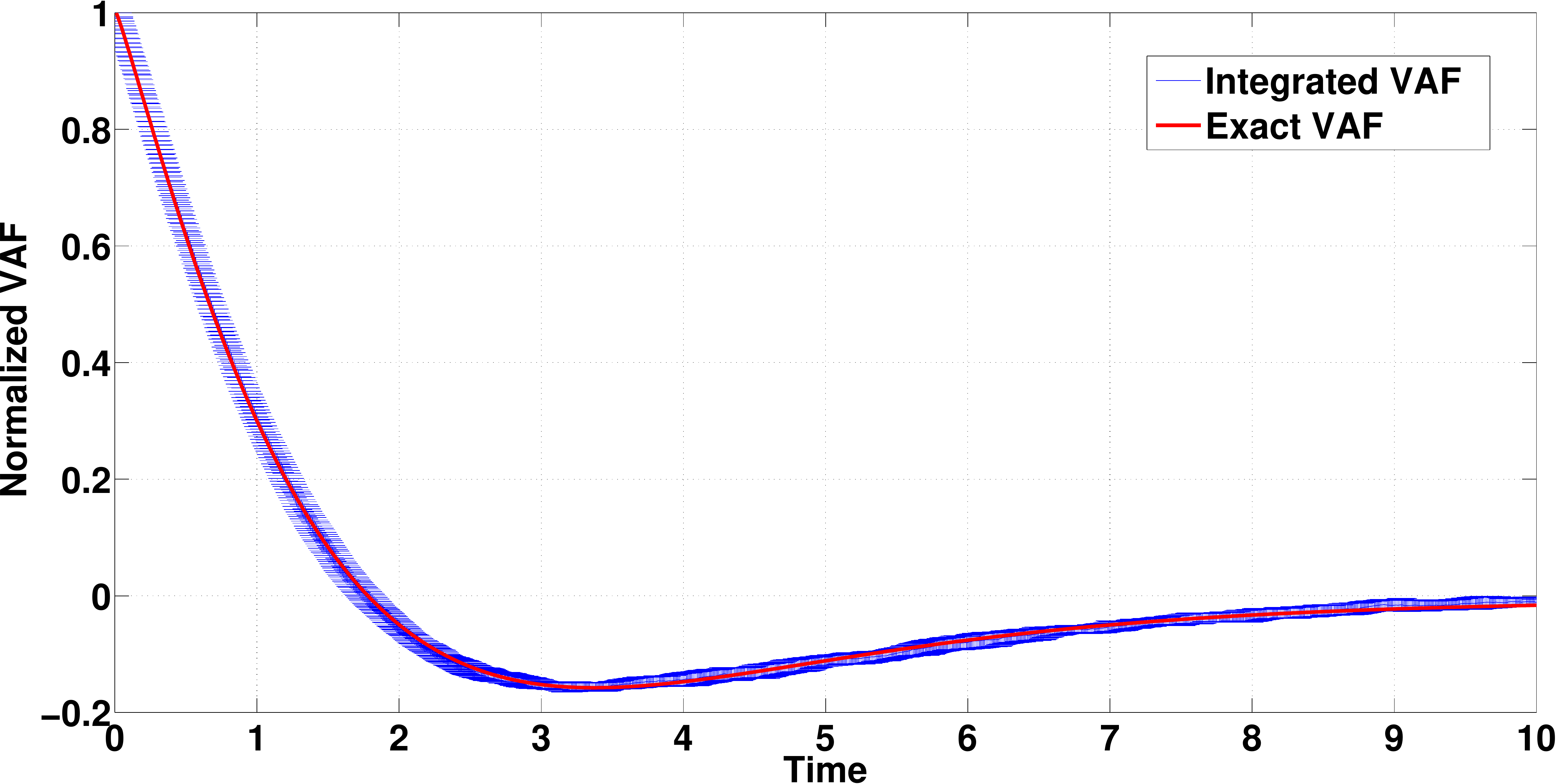}
 \caption{Illustration of the result of the kernel fitting procedure described in Section \ref{results}.  Note that while the exact VAF evolves under the influence of a harmonic confining potential (Equation \ref{harmonic_gle}), the integrated VAF is spawned by a GLE free of any conservative forces (Equation \ref{force_free_gle}).  The effect of the force is included in the construction of $\Gamma_{eff}(t)$. \label{harmonic_kernel_fit}}
\end{figure}

Here, the parameters utilized in Equation \ref{harmonic_gle} were $\omega_0=0.5$ 
and $\lambda=0.9$, and the associated analytic VAF was fit to a three term Prony series.  
This initial fit was done to realize a simpler form of the Laplace domain VAF, which 
was inserted into Equation \ref{kernel_fit_eqn}.  The resultant $\Gamma_{eff}(t)$
was also fit to a three term Prony series.  It is interesting to note that one 
term of this fit reduced to a conventional Langevin term (i.e., one that was proportional
to $\delta(t)$ rather than $\exp\left[-t/\tau\right]$) highlighting the utility of our 
method's reproduction of the Langevin limit.  This Prony series representation 
of $\Gamma_{eff}(t)$ was used in Equation \ref{force_free_gle}, which was numerically integrated 
to yield the integrated VAF result in Figure \ref{harmonic_kernel_fit}.

This result is especially interesting as it demonstrates capability for effecting a 
confining potential, on a finite timescale, exclusively through the GLE drag/random forces.  
This same fitting procedure could be used to remove inter-particle interactions, albeit
with some loss of dynamical information, and will be discussed in more detail
in future work. 

\begin{figure}[ht]
 \centering
 \graphicsmacro{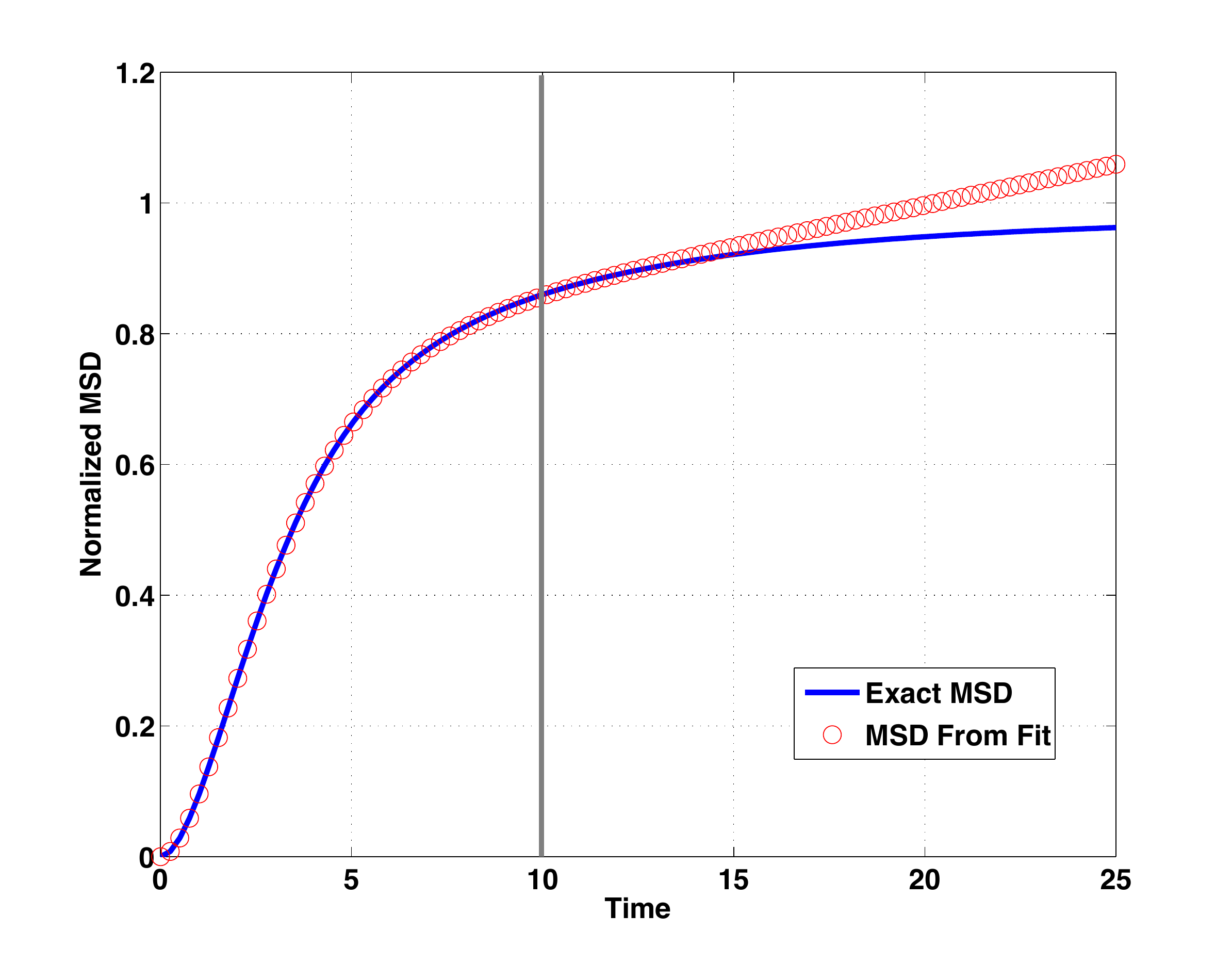}
 \caption{Comparison of the exact normalized MSD and the MSD generated from integrating the VAF fit.  The exact MSD is normalized such that its asymptotic limit is 1, and the same normalization is applied to the MSD integrated from the fit.  The grey line indicates the upper bound of the interval over which the VAF fit was constructed. \label{harmonic_msd_comparison}} 
\end{figure}

It is important to note that while the dynamics of a confined particle are 
reproduced in the absence of an explicit confining force, this is only the case over
the interval for which the memory kernel is reconstructed.  Outside of this interval,
the asymptotic behavior of the Prony series memory kernel will give rise to unbounded 
diffusive motion.  While the expected discrepancies in the dynamics are difficult to
notice in the VAF, as both the analytic and reconstructed quantities decay to zero, 
it is very evident in the mean-squared displacement (MSD). Here the asymptotic exponential 
behavior of the Prony series memory kernel necessarily leads to asymptotic diffusive motion
signaled by a linear MSD.  In contrast, the analytic memory kernel with the confining force 
will generate an MSD that remains bounded.  This is illustrated in Figure 
\ref{harmonic_msd_comparison}.  Here, the fit VAF is integrated to yield the MSD for
the Prony series model, and compared with the exact MSD over both the interval of the fit, 
and beyond.  This example illustrates the importance of choosing an appropriate interval 
for fitting memory kernels to achieve the appropriate limiting behavior in one's dynamics.

\section{Conclusions}
A family of numerical integration schemes for GLD have been presented.  These 
schemes are based upon an extended variable formulation of the GLE in which 
the memory kernel is rendered as a positive Prony series.  In certain limits,
it can be shown that a specific instance of this family of integrators 
exactly conserves the first and second moments of the integrated velocity 
distribution, and stably approaches the Langevin limit.  To this end, 
we identify this parametrization as optimal, and have implemented it in the MD 
code, LAMMPS.  Numerical experiments indicate that this implementation is 
robust for a number of canonical problems, as well as certain pathologies in 
the memory kernel.  An exemplary application to reduced order modeling 
illustrates potential uses of this module for MD practitioners.  
Future work will further develop the VAF fitting procedure in the context of 
statistical inference methods, and present extensions of the numerical 
integrator to mixed sign and complex memory kernels.

\section{Acknowledgements}
The authors would like to acknowledge Jason Bernstein, Paul Crozier, John Fricks, Jeremy Lechman, Rich Lehoucq, Scott McKinley, and Steve Plimpton for numerous fruitful discussions and feedback. Sandia National Laboratories is a multi-program laboratory managed and operated by Sandia Corporation, a wholly owned subsidiary of Lockheed Martin Corporation, for the U.S. Department of Energy’s National Nuclear Security Administration under contract DE-AC04-94AL85000.


\begin{thebibliography}{42}%
\makeatletter
\providecommand \@ifxundefined [1]{%
 \@ifx{#1\undefined}
}%
\providecommand \@ifnum [1]{%
 \ifnum #1\expandafter \@firstoftwo
 \else \expandafter \@secondoftwo
 \fi
}%
\providecommand \@ifx [1]{%
 \ifx #1\expandafter \@firstoftwo
 \else \expandafter \@secondoftwo
 \fi
}%
\providecommand \natexlab [1]{#1}%
\providecommand \enquote  [1]{``#1''}%
\providecommand \bibnamefont  [1]{#1}%
\providecommand \bibfnamefont [1]{#1}%
\providecommand \citenamefont [1]{#1}%
\providecommand \href@noop [0]{\@secondoftwo}%
\providecommand \href [0]{\begingroup \@sanitize@url \@href}%
\providecommand \@href[1]{\@@startlink{#1}\@@href}%
\providecommand \@@href[1]{\endgroup#1\@@endlink}%
\providecommand \@sanitize@url [0]{\catcode `\\12\catcode `\$12\catcode
  `\&12\catcode `\#12\catcode `\^12\catcode `\_12\catcode `\%12\relax}%
\providecommand \@@startlink[1]{}%
\providecommand \@@endlink[0]{}%
\providecommand \url  [0]{\begingroup\@sanitize@url \@url }%
\providecommand \@url [1]{\endgroup\@href {#1}{\urlprefix }}%
\providecommand \urlprefix  [0]{URL }%
\providecommand \Eprint [0]{\href }%
\providecommand \doibase [0]{http://dx.doi.org/}%
\providecommand \selectlanguage [0]{\@gobble}%
\providecommand \bibinfo  [0]{\@secondoftwo}%
\providecommand \bibfield  [0]{\@secondoftwo}%
\providecommand \translation [1]{[#1]}%
\providecommand \BibitemOpen [0]{}%
\providecommand \bibitemStop [0]{}%
\providecommand \bibitemNoStop [0]{.\EOS\space}%
\providecommand \EOS [0]{\spacefactor3000\relax}%
\providecommand \BibitemShut  [1]{\csname bibitem#1\endcsname}%
\let\auto@bib@innerbib\@empty
\bibitem [{\citenamefont {Coffey}, \citenamefont {Kalmykov},\ and\
  \citenamefont {Waldron}(2004)}]{coffey04}%
  \BibitemOpen
  \bibfield  {author} {\bibinfo {author} {\bibfnamefont {W.~T.}\ \bibnamefont
  {Coffey}}, \bibinfo {author} {\bibfnamefont {Y.~P.}\ \bibnamefont
  {Kalmykov}}, \ and\ \bibinfo {author} {\bibfnamefont {J.~T.}\ \bibnamefont
  {Waldron}},\ }\href@noop {} {\emph {\bibinfo {title} {The {L}angevin Equation
  With Applications to Stochastic Problems in Physics, Chemistry, and
  Electrical Engineering}}}\ (\bibinfo  {publisher} {World Scientific},\
  \bibinfo {year} {2004})\BibitemShut {NoStop}%
\bibitem [{\citenamefont {Uhlenbeck}\ and\ \citenamefont
  {Ornstein}(1930)}]{uhlenbeck30}%
  \BibitemOpen
  \bibfield  {author} {\bibinfo {author} {\bibfnamefont {G.~E.}\ \bibnamefont
  {Uhlenbeck}}\ and\ \bibinfo {author} {\bibfnamefont {L.~S.}\ \bibnamefont
  {Ornstein}},\ }\href@noop {} {\bibfield  {journal} {\bibinfo  {journal}
  {Phys. Rev.}\ }\textbf {\bibinfo {volume} {36}},\ \bibinfo {pages} {823}
  (\bibinfo {year} {1930})}\BibitemShut {NoStop}%
\bibitem [{\citenamefont {Kubo}(1966)}]{kubo66}%
  \BibitemOpen
  \bibfield  {author} {\bibinfo {author} {\bibfnamefont {R.}~\bibnamefont
  {Kubo}},\ }\href@noop {} {\bibfield  {journal} {\bibinfo  {journal} {Rep.
  Prog. Phys.}\ }\textbf {\bibinfo {volume} {29}},\ \bibinfo {pages} {255}
  (\bibinfo {year} {1966})}\BibitemShut {NoStop}%
\bibitem [{\citenamefont {Ermak}\ and\ \citenamefont
  {McCammon}(1978)}]{ermak78}%
  \BibitemOpen
  \bibfield  {author} {\bibinfo {author} {\bibfnamefont {D.~L.}\ \bibnamefont
  {Ermak}}\ and\ \bibinfo {author} {\bibfnamefont {J.~A.}\ \bibnamefont
  {McCammon}},\ }\href@noop {} {\bibfield  {journal} {\bibinfo  {journal} {J.
  Chem. Phys.}\ }\textbf {\bibinfo {volume} {69}},\ \bibinfo {pages} {1352}
  (\bibinfo {year} {1978})}\BibitemShut {NoStop}%
\bibitem [{\citenamefont {Schneider}\ and\ \citenamefont
  {Stoll}(1978)}]{schneider78}%
  \BibitemOpen
  \bibfield  {author} {\bibinfo {author} {\bibfnamefont {T.}~\bibnamefont
  {Schneider}}\ and\ \bibinfo {author} {\bibfnamefont {E.}~\bibnamefont
  {Stoll}},\ }\href@noop {} {\bibfield  {journal} {\bibinfo  {journal} {Phys.
  Rev. B}\ }\textbf {\bibinfo {volume} {17}},\ \bibinfo {pages} {1302}
  (\bibinfo {year} {1978})}\BibitemShut {NoStop}%
\bibitem [{\citenamefont {Izaguirre}\ \emph {et~al.}(2001)\citenamefont
  {Izaguirre}, \citenamefont {Caterello}, \citenamefont {Wozniak},\ and\
  \citenamefont {Skeel}}]{izaguirre01}%
  \BibitemOpen
  \bibfield  {author} {\bibinfo {author} {\bibfnamefont {J.~A.}\ \bibnamefont
  {Izaguirre}}, \bibinfo {author} {\bibfnamefont {D.~P.}\ \bibnamefont
  {Caterello}}, \bibinfo {author} {\bibfnamefont {J.~M.}\ \bibnamefont
  {Wozniak}}, \ and\ \bibinfo {author} {\bibfnamefont {R.~D.}\ \bibnamefont
  {Skeel}},\ }\href@noop {} {\bibfield  {journal} {\bibinfo  {journal} {J.
  Chem. Phys.}\ }\textbf {\bibinfo {volume} {114}},\ \bibinfo {pages} {2090}
  (\bibinfo {year} {2001})}\BibitemShut {NoStop}%
\bibitem [{\citenamefont {Mori}(1965{\natexlab{a}})}]{mori65}%
  \BibitemOpen
  \bibfield  {author} {\bibinfo {author} {\bibfnamefont {H.}~\bibnamefont
  {Mori}},\ }\href@noop {} {\bibfield  {journal} {\bibinfo  {journal} {Prog.
  Theo. Phys.}\ }\textbf {\bibinfo {volume} {33}},\ \bibinfo {pages} {423}
  (\bibinfo {year} {1965}{\natexlab{a}})}\BibitemShut {NoStop}%
\bibitem [{\citenamefont {Doll}\ and\ \citenamefont {Dion}(1976)}]{doll76}%
  \BibitemOpen
  \bibfield  {author} {\bibinfo {author} {\bibfnamefont {J.}~\bibnamefont
  {Doll}}\ and\ \bibinfo {author} {\bibfnamefont {D.}~\bibnamefont {Dion}},\
  }\href@noop {} {\bibfield  {journal} {\bibinfo  {journal} {J. Chem. Phys.}\
  }\textbf {\bibinfo {volume} {65}},\ \bibinfo {pages} {3762} (\bibinfo {year}
  {1976})}\BibitemShut {NoStop}%
\bibitem [{\citenamefont {Shugard}, \citenamefont {Tully},\ and\ \citenamefont
  {Nitzan}(1977)}]{shugard77}%
  \BibitemOpen
  \bibfield  {author} {\bibinfo {author} {\bibfnamefont {M.}~\bibnamefont
  {Shugard}}, \bibinfo {author} {\bibfnamefont {J.}~\bibnamefont {Tully}}, \
  and\ \bibinfo {author} {\bibfnamefont {A.}~\bibnamefont {Nitzan}},\
  }\href@noop {} {\bibfield  {journal} {\bibinfo  {journal} {J. Chem. Phys.}\
  }\textbf {\bibinfo {volume} {66}},\ \bibinfo {pages} {2534} (\bibinfo {year}
  {1977})}\BibitemShut {NoStop}%
\bibitem [{\citenamefont {Toxvaerd}(1987)}]{toxvaerd87}%
  \BibitemOpen
  \bibfield  {author} {\bibinfo {author} {\bibfnamefont {S.}~\bibnamefont
  {Toxvaerd}},\ }\href@noop {} {\bibfield  {journal} {\bibinfo  {journal} {J.
  Chem. Phys.}\ }\textbf {\bibinfo {volume} {86}},\ \bibinfo {pages} {3667}
  (\bibinfo {year} {1987})}\BibitemShut {NoStop}%
\bibitem [{\citenamefont {Schweizer}(1989)}]{schweizer89}%
  \BibitemOpen
  \bibfield  {author} {\bibinfo {author} {\bibfnamefont {K.}~\bibnamefont
  {Schweizer}},\ }\href@noop {} {\bibfield  {journal} {\bibinfo  {journal} {J.
  Chem. Phys.}\ }\textbf {\bibinfo {volume} {91}},\ \bibinfo {pages} {5802}
  (\bibinfo {year} {1989})}\BibitemShut {NoStop}%
\bibitem [{\citenamefont {Mason}\ and\ \citenamefont {Weitz}(1995)}]{mason95}%
  \BibitemOpen
  \bibfield  {author} {\bibinfo {author} {\bibfnamefont {T.~G.}\ \bibnamefont
  {Mason}}\ and\ \bibinfo {author} {\bibfnamefont {D.~A.}\ \bibnamefont
  {Weitz}},\ }\href@noop {} {\bibfield  {journal} {\bibinfo  {journal} {Phys.
  Rev. Lett.}\ }\textbf {\bibinfo {volume} {74}},\ \bibinfo {pages} {1250}
  (\bibinfo {year} {1995})}\BibitemShut {NoStop}%
\bibitem [{\citenamefont {Mason}, \citenamefont {Gang},\ and\ \citenamefont
  {Weitz}(1997)}]{mason97}%
  \BibitemOpen
  \bibfield  {author} {\bibinfo {author} {\bibfnamefont {T.~G.}\ \bibnamefont
  {Mason}}, \bibinfo {author} {\bibfnamefont {H.}~\bibnamefont {Gang}}, \ and\
  \bibinfo {author} {\bibfnamefont {D.~A.}\ \bibnamefont {Weitz}},\ }\href@noop
  {} {\bibfield  {journal} {\bibinfo  {journal} {J. Opt. Soc. Am. A}\ }\textbf
  {\bibinfo {volume} {14}},\ \bibinfo {pages} {139} (\bibinfo {year}
  {1997})}\BibitemShut {NoStop}%
\bibitem [{\citenamefont {Fricks}\ \emph {et~al.}(2009)\citenamefont {Fricks},
  \citenamefont {Yao}, \citenamefont {Elston},\ and\ \citenamefont
  {Forest}}]{fricks09}%
  \BibitemOpen
  \bibfield  {author} {\bibinfo {author} {\bibfnamefont {J.}~\bibnamefont
  {Fricks}}, \bibinfo {author} {\bibfnamefont {L.}~\bibnamefont {Yao}},
  \bibinfo {author} {\bibfnamefont {T.~C.}\ \bibnamefont {Elston}}, \ and\
  \bibinfo {author} {\bibfnamefont {M.~G.}\ \bibnamefont {Forest}},\
  }\href@noop {} {\bibfield  {journal} {\bibinfo  {journal} {SIAM J. Appl.
  Math.}\ }\textbf {\bibinfo {volume} {69}},\ \bibinfo {pages} {1277} (\bibinfo
  {year} {2009})}\BibitemShut {NoStop}%
\bibitem [{\citenamefont {Didier}\ \emph {et~al.}(2012)\citenamefont {Didier},
  \citenamefont {McKinley}, \citenamefont {Hill},\ and\ \citenamefont
  {Fricks}}]{didier12}%
  \BibitemOpen
  \bibfield  {author} {\bibinfo {author} {\bibfnamefont {G.}~\bibnamefont
  {Didier}}, \bibinfo {author} {\bibfnamefont {S.~A.}\ \bibnamefont
  {McKinley}}, \bibinfo {author} {\bibfnamefont {D.~B.}\ \bibnamefont {Hill}},
  \ and\ \bibinfo {author} {\bibfnamefont {J.}~\bibnamefont {Fricks}},\
  }\href@noop {} {\bibfield  {journal} {\bibinfo  {journal} {J. Time Series
  Analysis}\ }\textbf {\bibinfo {volume} {33}},\ \bibinfo {pages} {724}
  (\bibinfo {year} {2012})}\BibitemShut {NoStop}%
\bibitem [{\citenamefont {Kou}\ and\ \citenamefont {Xie}(2004)}]{kou04}%
  \BibitemOpen
  \bibfield  {author} {\bibinfo {author} {\bibfnamefont {S.~C.}\ \bibnamefont
  {Kou}}\ and\ \bibinfo {author} {\bibfnamefont {X.~S.}\ \bibnamefont {Xie}},\
  }\href@noop {} {\bibfield  {journal} {\bibinfo  {journal} {Phys. Rev. Lett.}\
  }\textbf {\bibinfo {volume} {93}},\ \bibinfo {pages} {180603} (\bibinfo
  {year} {2004})}\BibitemShut {NoStop}%
\bibitem [{\citenamefont {Min}\ \emph {et~al.}(2005)\citenamefont {Min},
  \citenamefont {Luo}, \citenamefont {Cherayil}, \citenamefont {Kou},\ and\
  \citenamefont {Xie}}]{min05}%
  \BibitemOpen
  \bibfield  {author} {\bibinfo {author} {\bibfnamefont {W.}~\bibnamefont
  {Min}}, \bibinfo {author} {\bibfnamefont {G.}~\bibnamefont {Luo}}, \bibinfo
  {author} {\bibfnamefont {B.~J.}\ \bibnamefont {Cherayil}}, \bibinfo {author}
  {\bibfnamefont {S.~C.}\ \bibnamefont {Kou}}, \ and\ \bibinfo {author}
  {\bibfnamefont {X.~S.}\ \bibnamefont {Xie}},\ }\href@noop {} {\bibfield
  {journal} {\bibinfo  {journal} {Phys. Rev. Lett.}\ }\textbf {\bibinfo
  {volume} {94}},\ \bibinfo {pages} {198302} (\bibinfo {year}
  {2005})}\BibitemShut {NoStop}%
\bibitem [{\citenamefont {Gordon}, \citenamefont {Krishnamurthy},\ and\
  \citenamefont {Chung}(2009)}]{gordon09}%
  \BibitemOpen
  \bibfield  {author} {\bibinfo {author} {\bibfnamefont {D.}~\bibnamefont
  {Gordon}}, \bibinfo {author} {\bibfnamefont {V.}~\bibnamefont
  {Krishnamurthy}}, \ and\ \bibinfo {author} {\bibfnamefont {S.}~\bibnamefont
  {Chung}},\ }\href@noop {} {\bibfield  {journal} {\bibinfo  {journal} {J.
  Chem. Phys.}\ }\textbf {\bibinfo {volume} {131}},\ \bibinfo {pages} {134102}
  (\bibinfo {year} {2009})}\BibitemShut {NoStop}%
\bibitem [{\citenamefont {Ceriotti}, \citenamefont {Bussi},\ and\
  \citenamefont {Parrinello}(2009)}]{ceriotti09a}%
  \BibitemOpen
  \bibfield  {author} {\bibinfo {author} {\bibfnamefont {M.}~\bibnamefont
  {Ceriotti}}, \bibinfo {author} {\bibfnamefont {G.}~\bibnamefont
  {Bussi}}, \ and\ \bibinfo {author} {\bibfnamefont {M.}~\bibnamefont
  {Parrinello}},\ }\href@noop {} {\bibfield  {journal} {\bibinfo  {journal} {Phys.
  Rev. Lett.}\ }\textbf {\bibinfo {volume} {102}},\ \bibinfo {pages} {020601}
  (\bibinfo {year} {2009})}\BibitemShut {NoStop}%
\bibitem [{\citenamefont {Ceriotti}, \citenamefont {Bussi},\ and\
  \citenamefont {Parrinello}(2009)}]{ceriotti09b}%
  \BibitemOpen
  \bibfield  {author} {\bibinfo {author} {\bibfnamefont {M.}~\bibnamefont
  {Ceriotti}}, \bibinfo {author} {\bibfnamefont {G.}~\bibnamefont
  {Bussi}}, \ and\ \bibinfo {author} {\bibfnamefont {M.}~\bibnamefont
  {Parrinello}},\ }\href@noop {} {\bibfield  {journal} {\bibinfo  {journal} {Phys.
  Rev. Lett.}\ }\textbf {\bibinfo {volume} {103}},\ \bibinfo {pages} {030603}
  (\bibinfo {year} {2009})}\BibitemShut {NoStop}%
\bibitem [{\citenamefont {Ceriotti} and\
  \citenamefont {Manolopoulos}(2012)}]{ceriotti12}%
  \BibitemOpen
  \bibfield  {author} {\bibinfo {author} {\bibfnamefont {M.}~\bibnamefont
  {Ceriotti}} \ and\ \bibinfo {author} {\bibfnamefont {D.E..}~\bibnamefont
  {Manolopoulos}},\ }\href@noop {} {\bibfield  {journal} {\bibinfo  {journal} {Phys.
  Rev. Lett.}\ }\textbf {\bibinfo {volume} {109}},\ \bibinfo {pages} {100604}
  (\bibinfo {year} {2012})}\BibitemShut {NoStop}%
\bibitem [{\citenamefont {Sokolov}(2012)}]{sokolov12}%
  \BibitemOpen
  \bibfield  {author} {\bibinfo {author} {\bibfnamefont {I.}~\bibnamefont
  {Sokolov}},\ }\href@noop {} {\bibfield  {journal} {\bibinfo  {journal} {Soft
  Matter}\ }\textbf {\bibinfo {volume} {8}},\ \bibinfo {pages} {9043} (\bibinfo
  {year} {2012})}\BibitemShut {NoStop}%
\bibitem [{\citenamefont {Ciccotti}\ and\ \citenamefont
  {Ryckaert}(1980)}]{ciccotti80}%
  \BibitemOpen
  \bibfield  {author} {\bibinfo {author} {\bibfnamefont {G.}~\bibnamefont
  {Ciccotti}}\ and\ \bibinfo {author} {\bibfnamefont {J.}~\bibnamefont
  {Ryckaert}},\ }\href@noop {} {\bibfield  {journal} {\bibinfo  {journal} {Mol.
  Phys.}\ }\textbf {\bibinfo {volume} {40}},\ \bibinfo {pages} {141} (\bibinfo
  {year} {1980})}\BibitemShut {NoStop}%
\bibitem [{\citenamefont {Berkowitz}, \citenamefont {Morgan},\ and\
  \citenamefont {McCammon}(1983)}]{berkowitz83}%
  \BibitemOpen
  \bibfield  {author} {\bibinfo {author} {\bibfnamefont {M.}~\bibnamefont
  {Berkowitz}}, \bibinfo {author} {\bibfnamefont {J.}~\bibnamefont {Morgan}}, \
  and\ \bibinfo {author} {\bibfnamefont {J.}~\bibnamefont {McCammon}},\
  }\href@noop {} {\bibfield  {journal} {\bibinfo  {journal} {J. Chem. Phys.}\
  }\textbf {\bibinfo {volume} {78}},\ \bibinfo {pages} {3256} (\bibinfo {year}
  {1983})}\BibitemShut {NoStop}%
\bibitem [{\citenamefont {Gu{\`a}rdia}\ and\ \citenamefont
  {Padr{\'o}}(1985)}]{guardia85}%
  \BibitemOpen
  \bibfield  {author} {\bibinfo {author} {\bibfnamefont {E.}~\bibnamefont
  {Gu{\`a}rdia}}\ and\ \bibinfo {author} {\bibfnamefont {J.}~\bibnamefont
  {Padr{\'o}}},\ }\href@noop {} {\bibfield  {journal} {\bibinfo  {journal} {J.
  Chem. Phys.}\ }\textbf {\bibinfo {volume} {83}},\ \bibinfo {pages} {1917}
  (\bibinfo {year} {1985})}\BibitemShut {NoStop}%
\bibitem [{\citenamefont {Straub}, \citenamefont {Borkovec},\ and\
  \citenamefont {Berne}(1986)}]{straub86}%
  \BibitemOpen
  \bibfield  {author} {\bibinfo {author} {\bibfnamefont {J.}~\bibnamefont
  {Straub}}, \bibinfo {author} {\bibfnamefont {M.}~\bibnamefont {Borkovec}}, \
  and\ \bibinfo {author} {\bibfnamefont {B.}~\bibnamefont {Berne}},\
  }\href@noop {} {\bibfield  {journal} {\bibinfo  {journal} {J. Chem. Phys.}\
  }\textbf {\bibinfo {volume} {84}},\ \bibinfo {pages} {1788} (\bibinfo {year}
  {1986})}\BibitemShut {NoStop}%
\bibitem [{\citenamefont {Smith}\ and\ \citenamefont {Harris}(1990)}]{smith90}%
  \BibitemOpen
  \bibfield  {author} {\bibinfo {author} {\bibfnamefont {D.}~\bibnamefont
  {Smith}}\ and\ \bibinfo {author} {\bibfnamefont {C.}~\bibnamefont {Harris}},\
  }\href@noop {} {\bibfield  {journal} {\bibinfo  {journal} {J. Chem. Phys.}\
  }\textbf {\bibinfo {volume} {92}},\ \bibinfo {pages} {1304} (\bibinfo {year}
  {1990})}\BibitemShut {NoStop}%
\bibitem [{\citenamefont {Nilsson}\ and\ \citenamefont
  {Padr{\'o}}(1990)}]{nilsson90}%
  \BibitemOpen
  \bibfield  {author} {\bibinfo {author} {\bibfnamefont {L.}~\bibnamefont
  {Nilsson}}\ and\ \bibinfo {author} {\bibfnamefont {J.}~\bibnamefont
  {Padr{\'o}}},\ }\href@noop {} {\bibfield  {journal} {\bibinfo  {journal}
  {Mol. Phys.}\ }\textbf {\bibinfo {volume} {71}},\ \bibinfo {pages} {355}
  (\bibinfo {year} {1990})}\BibitemShut {NoStop}%
\bibitem [{\citenamefont {Wan}, \citenamefont {Wang},\ and\ \citenamefont
  {Shi}(1998)}]{wan98}%
  \BibitemOpen
  \bibfield  {author} {\bibinfo {author} {\bibfnamefont {S.}~\bibnamefont
  {Wan}}, \bibinfo {author} {\bibfnamefont {C.}~\bibnamefont {Wang}}, \ and\
  \bibinfo {author} {\bibfnamefont {Y.}~\bibnamefont {Shi}},\ }\href@noop {}
  {\bibfield  {journal} {\bibinfo  {journal} {Mol. Phys.}\ }\textbf {\bibinfo
  {volume} {93}},\ \bibinfo {pages} {901} (\bibinfo {year} {1998})}\BibitemShut
  {NoStop}%
\bibitem [{\citenamefont {Gordon}, \citenamefont {Krishnamurthy},\ and\
  \citenamefont {Chung}(2008)}]{gordon08}%
  \BibitemOpen
  \bibfield  {author} {\bibinfo {author} {\bibfnamefont {D.}~\bibnamefont
  {Gordon}}, \bibinfo {author} {\bibfnamefont {V.}~\bibnamefont
  {Krishnamurthy}}, \ and\ \bibinfo {author} {\bibfnamefont {S.}~\bibnamefont
  {Chung}},\ }\href@noop {} {\bibfield  {journal} {\bibinfo  {journal} {Mol.
  Phys.}\ }\textbf {\bibinfo {volume} {106}},\ \bibinfo {pages} {1353}
  (\bibinfo {year} {2008})}\BibitemShut {NoStop}%
\bibitem [{\citenamefont {Plimpton}(1995)}]{plimpton95}%
  \BibitemOpen
  \bibfield  {author} {\bibinfo {author} {\bibfnamefont {S.}~\bibnamefont
  {Plimpton}},\ }\href@noop {} {\bibfield  {journal} {\bibinfo  {journal} {J.
  Comp. Phys.}\ }\textbf {\bibinfo {volume} {117}},\ \bibinfo {pages} {1}
  (\bibinfo {year} {1995})},\ \bibinfo {note} {available at
  http://lammps.sandia.gov}\BibitemShut {NoStop}%
\bibitem [{\citenamefont {in 't Veld}(2008)}]{intveld08}%
  \BibitemOpen
  \bibfield  {author} {\bibinfo {author} {\bibfnamefont {P.}~\bibnamefont
  {in 't Veld}}, \bibinfo {author} {\bibfnamefont {S.J.}~\bibnamefont {Plimpton}}, \ and\
  \bibinfo {author} {\bibfnamefont {G.S.}~\bibnamefont {Grest}},\ }\href@noop {} {\bibfield  {journal} {\bibinfo {journal} {Comp.
  Phys. Comm.}\ }\textbf {\bibinfo {volume} {179}},\ \bibinfo {pages} {320}
  (\bibinfo {year} {2008})}\BibitemShut {NoStop}%
\bibitem [{\citenamefont {Desp{\'o}sito}\ and\ \citenamefont
  {Vi{\~n}ales}(2009)}]{desposito09}%
  \BibitemOpen
  \bibfield  {author} {\bibinfo {author} {\bibfnamefont {M.~A.}\ \bibnamefont
  {Desp{\'o}sito}}\ and\ \bibinfo {author} {\bibfnamefont {A.~D.}\ \bibnamefont
  {Vi{\~n}ales}},\ }\href@noop {} {\bibfield  {journal} {\bibinfo  {journal}
  {Phys. Rev. E}\ }\textbf {\bibinfo {volume} {80}},\ \bibinfo {pages} {021111}
  (\bibinfo {year} {2009})}\BibitemShut {NoStop}%
\bibitem [{\citenamefont {Vi{\~n}ales}\ and\ \citenamefont
  {Desp\'osito}(2006)}]{vinales06}%
  \BibitemOpen
  \bibfield  {author} {\bibinfo {author} {\bibfnamefont {A.~D.}\ \bibnamefont
  {Vi{\~n}ales}}\ and\ \bibinfo {author} {\bibfnamefont {M.~A.}\ \bibnamefont
  {Desp\'osito}},\ }\href@noop {} {\bibfield  {journal} {\bibinfo  {journal}
  {Phys. Rev. E}\ }\textbf {\bibinfo {volume} {73}},\ \bibinfo {pages} {016111}
  (\bibinfo {year} {2006})}\BibitemShut {NoStop}%
\bibitem [{\citenamefont {McKinley}, \citenamefont {Yao},\ and\ \citenamefont
  {Gregory}(2009)}]{mckinley09}%
  \BibitemOpen
  \bibfield  {author} {\bibinfo {author} {\bibfnamefont {S.}~\bibnamefont
  {McKinley}}, \bibinfo {author} {\bibfnamefont {L.}~\bibnamefont {Yao}}, \
  and\ \bibinfo {author} {\bibfnamefont {F.}~\bibnamefont {Gregory}},\
  }\href@noop {} {\bibfield  {journal} {\bibinfo  {journal} {J. Rheology}\
  }\textbf {\bibinfo {volume} {53}},\ \bibinfo {pages} {1487} (\bibinfo {year}
  {2009})}\BibitemShut {NoStop}%
\bibitem [{\citenamefont {Ottobre}\ and\ \citenamefont
  {Pavliotis}(2011)}]{ottobre11}%
  \BibitemOpen
  \bibfield  {author} {\bibinfo {author} {\bibfnamefont {M.}~\bibnamefont
  {Ottobre}}\ and\ \bibinfo {author} {\bibfnamefont {G.~A.}\ \bibnamefont
  {Pavliotis}},\ }\href@noop {} {\bibfield  {journal} {\bibinfo  {journal}
  {Nonlinearity}\ }\textbf {\bibinfo {volume} {24}},\ \bibinfo {pages} {1629}
  (\bibinfo {year} {2011})}\BibitemShut {NoStop}%
\bibitem [{\citenamefont {Wang}\ and\ \citenamefont {Skeel}(2003)}]{wang03}%
  \BibitemOpen
  \bibfield  {author} {\bibinfo {author} {\bibfnamefont {W.}~\bibnamefont
  {Wang}}\ and\ \bibinfo {author} {\bibfnamefont {R.~D.}\ \bibnamefont
  {Skeel}},\ }\href@noop {} {\bibfield  {journal} {\bibinfo  {journal}
  {Molecular Physics}\ }\textbf {\bibinfo {volume} {101}},\ \bibinfo {pages}
  {2149} (\bibinfo {year} {2003})}\BibitemShut {NoStop}%
\bibitem [{\citenamefont {Callen}\ and\ \citenamefont
  {Welton}(1951)}]{callen51}%
  \BibitemOpen
  \bibfield  {author} {\bibinfo {author} {\bibfnamefont {H.~B.}\ \bibnamefont
  {Callen}}\ and\ \bibinfo {author} {\bibfnamefont {T.~A.}\ \bibnamefont
  {Welton}},\ }\href@noop {} {\bibfield  {journal} {\bibinfo  {journal} {Phys.
  Rev}\ }\textbf {\bibinfo {volume} {83}},\ \bibinfo {pages} {34} (\bibinfo
  {year} {1951})}\BibitemShut {NoStop}%
\bibitem [{\citenamefont {Mori}(1965{\natexlab{b}})}]{mori65_2}%
  \BibitemOpen
  \bibfield  {author} {\bibinfo {author} {\bibfnamefont {H.}~\bibnamefont
  {Mori}},\ }\href@noop {} {\bibfield  {journal} {\bibinfo  {journal} {Prog.
  Theo. Phys.}\ }\textbf {\bibinfo {volume} {34}},\ \bibinfo {pages} {399}
  (\bibinfo {year} {1965}{\natexlab{b}})}\BibitemShut {NoStop}%
\bibitem [{\citenamefont {Kupferman}(2004)}]{kupferman04}%
  \BibitemOpen
  \bibfield  {author} {\bibinfo {author} {\bibfnamefont {R.}~\bibnamefont
  {Kupferman}},\ }\href@noop {} {\bibfield  {journal} {\bibinfo  {journal} {J.
  Stat. Phys.}\ }\textbf {\bibinfo {volume} {114}},\ \bibinfo {pages} {291}
  (\bibinfo {year} {2004})}\BibitemShut {NoStop}%
\bibitem [{\citenamefont {Kloeden}\ and\ \citenamefont
  {Platen}(1999)}]{kloeden_book}%
  \BibitemOpen
  \bibfield  {author} {\bibinfo {author} {\bibfnamefont {P.~E.}\ \bibnamefont
  {Kloeden}}\ and\ \bibinfo {author} {\bibfnamefont {E.}~\bibnamefont
  {Platen}},\ }\href@noop {} {\emph {\bibinfo {title} {Numerical Solution of
  Stochastic Differential Equations}}}\ (\bibinfo  {publisher} {Springer},\
  \bibinfo {year} {1999})\BibitemShut {NoStop}%
\bibitem [{\citenamefont {Hairer}, \citenamefont {Lubich},\ and\ \citenamefont
  {Wanner}(2003)}]{hairer03}%
  \BibitemOpen
  \bibfield  {author} {\bibinfo {author} {\bibfnamefont {E.}~\bibnamefont
  {Hairer}}, \bibinfo {author} {\bibfnamefont {C.}~\bibnamefont {Lubich}}, \
  and\ \bibinfo {author} {\bibfnamefont {G.}~\bibnamefont {Wanner}},\
  }\href@noop {} {\bibfield  {journal} {\bibinfo  {journal} {Acta Numerica}\
  }\textbf {\bibinfo {volume} {12}},\ \bibinfo {pages} {399} (\bibinfo {year}
  {2003})}\BibitemShut {NoStop}%
\bibitem [{\citenamefont {Leimkuhler}\ and\ \citenamefont
  {Matthews}(2013)}]{leimkuhler13}%
  \BibitemOpen
  \bibfield  {author} {\bibinfo {author} {\bibfnamefont {B.}~\bibnamefont
  {Leimkuhler}}\ and\ \bibinfo {author} {\bibfnamefont {C.}~\bibnamefont
  {Matthews}},\ }\href@noop {} {\bibfield  {journal} {\bibinfo  {journal}
  {Applied Mathematics Research eXpress}\ }\textbf {\bibinfo {volume} {2013}},\
  \bibinfo {pages} {34} (\bibinfo {year} {2013})}\BibitemShut {NoStop}%
\bibitem [{\citenamefont {D\"unweg}\ and\ \citenamefont
  {Paul}(1991)}]{dunweg91}%
  \BibitemOpen
  \bibfield  {author} {\bibinfo {author} {\bibfnamefont {B.}~\bibnamefont
  {D\"unweg}}\ and\ \bibinfo {author} {\bibfnamefont {W.}~\bibnamefont
  {Paul}},\ }\href@noop {} {\bibfield  {journal} {\bibinfo  {journal}
  {International Journal of Modern Physics C}\ }\textbf {\bibinfo {volume}
  {2}},\ \bibinfo {pages} {817} (\bibinfo {year} {1991})}\BibitemShut {NoStop}%
\bibitem [{\citenamefont {Allen}\ and\ \citenamefont
  {Tildesley}(1987)}]{allenbook}%
  \BibitemOpen
  \bibfield  {author} {\bibinfo {author} {\bibfnamefont {M.}~\bibnamefont
  {Allen}}\ and\ \bibinfo {author} {\bibfnamefont {D.}~\bibnamefont
  {Tildesley}},\ }\href@noop {} {\emph {\bibinfo {title} {Computer Simulation
  of Liquids}}}\ (\bibinfo  {publisher} {Oxford University Press},\ \bibinfo
  {year} {1987})\BibitemShut {NoStop}%
\bibitem [{\citenamefont {Berne}, \citenamefont {Boon},\ and\ \citenamefont
  {Rice}(1966)}]{berne66}%
  \BibitemOpen
  \bibfield  {author} {\bibinfo {author} {\bibfnamefont {B.~J.}\ \bibnamefont
  {Berne}}, \bibinfo {author} {\bibfnamefont {J.~P.}\ \bibnamefont {Boon}}, \
  and\ \bibinfo {author} {\bibfnamefont {S.~A.}\ \bibnamefont {Rice}},\
  }\href@noop {} {\bibfield  {journal} {\bibinfo  {journal} {J. Chem. Phys.}\
  }\textbf {\bibinfo {volume} {45}},\ \bibinfo {pages} {1086} (\bibinfo {year}
  {1966})}\BibitemShut {NoStop}%
\end{thebibliography}

%

\end{document}